\documentclass[%
floatfix,
 aapm,
 mph,%
 amsmath,amssymb,
 reprint%
]{revtex4-2}

\usepackage{graphicx}
\usepackage{dcolumn}
\usepackage{bm}
\usepackage{xcolor}
\usepackage[mathlines]{lineno}
\modulolinenumbers[5]

\begin{document}

\preprint{AAPM/123-QED}

\title{Simulations of Collisional Effects in an Inner-Shell Solid-Density Mg X-Ray Laser}

\author{Shenyuan Ren}
\author{Sam Vinko}
\author{Justin S. Wark}
 \address{Department of Physics, Clarendon Laboratory, University of Oxford, Parks Road, Oxford, OX1 3PU, United Kingdom}

\date{\today}

\begin{abstract}
Inner-shell K$\alpha$ x-ray lasers have been created by pumping gaseous, solid, and liquid targets with the intense x-ray output of free-electron-lasers (FELs). For gaseous targets lasing relies on the creation of K-shell core-holes on a time-scale short compared with filling via Auger decay. In the case of solid and liquid density systems, collisional effects will also be important, affecting not only populations, but also line-widths, both of which impact the degree of overall gain, and its duration.  However,  to date such collisional effects have not been extensively studied. We present here initial simulations using the CCFLY code of inner-shell lasing in solid density Mg, where we self-consistently treat the effects of the incoming FEL radiation and the atomic kinetics of the Mg system, including radiative, Auger, and collisional effects. We find that the combination of collisional population of the lower states of the lasing transitions and broadening of the lines precludes lasing on all but the K$\alpha$ of the initially cold system.  Even assuming instantaneous turning on of the FEL pump, we find the duration of the gain in the solid system to be sub-femtosecond.
\end{abstract}

\maketitle

\section{\label{sec:intro}Introduction\protect\\ }

The concept of producing short-wavelength lasers via the photoionization of inner-shell electrons was put forward more than half a century ago, when detailed calculations of the requirements for lasing on the K-shell of Sodium and Copper were performed~\cite{Duguay1967}.  Owing to the large power-densities and short rise times required for pumping, the practical realisation of such ideas took some considerable time to be realised in the x-ray regime~\cite{Rohringer2012,Yoneda2015,Kroll2018}, although lasing in the XUV was successfully demonstrated with other schemes based on collisional excitation~\cite{Matthews1985} or recombination~\cite{Popovics1987} in highly ionized plasmas~\cite{Matthews1995,Jaegle_2006}.

The photoionization approach was first successfully demonstrated in the optical regime (0.325-$\mu$m) where laser-plasma-produced soft x-rays were used to photoeject a $d$-shell electron from neutral Cd~\cite{Silfvast1983}. A similar concept with the higher $Z$ material Xenon allowed for an extension into the soft x-ray regime at 0.1089-$\mu$m~\cite{Kapteyn1986}. As noted above, extension to the keV photon-energy regime is much more difficult, owing to the non-linear scaling of the required pump power with the inverse of the wavelength of the lasing transition. For this reason successful demonstration of a K-shell x-ray laser~\cite{Rohringer2012,Yoneda2015,Kroll2018} had to await the development of the high brightness x-ray pump sources provided by x-ray FELs~\cite{Pellegrini2012}.

The requirement for high pump power is related to the short lifetime of the relevant transition, whereby the lifetime of a K-shell hole created in an isolated neutral atom or ion will normally be dominated by the Auger decay rate.  For mid-$Z$ elements the Auger lifetime is typically in the femtosecond regime. Whilst in a low density system the Auger rate will be the self-terminating factor, at higher densities electron collisional ionisation becomes important. As pointed out by Kapteyn~\cite{Kapteyn1992} in a dense system the electrons that are ejected into the continuum due to photoionisation of the K-shell and the subsequent Auger decay processes will, via collisional processes,  produce copious higher ionisation stages in their ground state, which comprise the lower laser level of the K-shell transitions, thus potentially curtailing gain.  For a system of sufficiently low density, such collisional effects should not be important, and the rate of pumping is determined simply by the requirement to overcome the Auger decay rate.  Detailed simulations by Rohringer {\it et al.} indicated that the x-ray intensities afforded by hard x-ray FEL systems would be capable of achieving  gain in a Neon gas~\cite{Rohringer2007,Rohringer2009}; a prediction that was subsequently verified experimentally by the same lead author a few years later in the first successful demonstration of a K-shell x-ray laser.  In that experiment, performed at the Linac Coherent Light Source (LCLS), gaseous Neon was irradiated by 0.96 keV x-rays at an intensity of order $2 \times 10^{17}$ W cm$^{-2}$ in pulses between 40 and 80 fs in duration. Lasing was observed on the K$\alpha$ transition of neutral Neon at 0.849 keV~\cite{Rohringer2012}.

Whilst K-shell lasing in a solid density system is in principle more difficult owing to the collisional effects mentioned above, such have been the advances in FEL technology that a few years later gain was observed on the K$\alpha 1$ and K$\alpha 2$ transitions of Cu at 8.04 and 8.02 keV respectively.  In these experiments, performed at the SACLA BL3 facility~\cite{Tono2013} the FEL output, of 7 fs duration, was focussed onto solid Cu targets at irradiances of several times $10^{19}$ W cm$^{-2}$. The authors also used a two-colour FEL scheme to seed the lasing transition~\cite{Yoneda2015}.  Further developments have been made in lasing in liquid systems.  In experiments at LCLS stimulated emission has been observed on the K$\alpha 1$ transition of Mn by the irradiation of two different Mn compounds, Mn(II)Cl$_2$ and NaMn(VII)O$_4$, in aqueous solutions.  In these experiments the LCLS beam was tuned to an energy of 6.6 keV, with a pulse length estimated to be between 10 and 30 fs at an intensity of order 10$^{20}$W cm$^{-2}$.  Detailed measurements of the lasing output provided estimates of a gain of order 2 $\times$ 10$^6$ over the spontaneous emission~\cite{Kroll2018}, with an assumed output pulse duration of order 1 fs. Interestingly, the photon energy of the K$\alpha 1$ emission was found to be slightly different for the two solutions, indicating that the effect of the chemical environment was preserved during the lasing process, hinting that significant gain existed before any outer-shell ionisation had occurred.

The work cited above demonstrates the interest in generating K-shell x-ray lasers using FELs as the pump, and it is in this context that we present here our initial simulations of K-shell lasing in both low density and solid-density Mg.  This study is in part motivated by the observation that most simulations or calculations to date of FEL pumped K-shell lasing have been for low density systems~\cite{Rohringer2007,Rohringer2009,Yoneda2015,Nilsen2016}, rather than explicitly taking collisional effects into account.  

We choose Mg for a number of reasons.  Firstly, as a relatively low-Z material the atomic physics is less complex, reducing computational cost.  Secondly, there have been several experimental studies of the spontaneous emission spectra emitted by solid density Mg targets illuminated by an FEL, and comparison of those spectra with previous simulations with the CCFLY code used here
 provides confidence that the overall evolution of the charge state distribution in the solid-density plasma is relatively well-understood~\cite{Ciricosta2016a, Preston2017, Vandenberg2018}.
Lastly, with an atomic number of 12, it does not differ too much from that of Neon (Z=10), allowing at least some degree of comparison of results (at least in the low density case) with those of Rohringer, although the details of the calculations differ in several respects which we outline below.  

\section{Simulations}

Simulations were performed using the collisional-radiative code CCFLY, which itself is an updated (written in c++) version of the SCFLY code described briefly elsewhere~\cite{Ciricosta2016}, with both these codes being substantially revised versions of the widely available FLYCHK suite~\cite{Chung2005,Chung2007}.  CCFLY is a non-LTE code in that the emitted radiation and ionic ground and excited-state populations are evolved in time, rather than assuming thermodynamic equilibrium.  Note, however, that in the version of the code used here, electrons in the continuum are assumed to obey classical statistics, and to instantaneously thermalise to a temperature dictated by their overall energy content.  Within that caveat CCFLY is specifically tailored for X-ray laser problems in that it provides a self consistent electron temperature calculation derived by the energy balance between the absorbed FEL radiation, internal energies of the electrons and ions in the system, and the emitted radiation.  Whilst energy can be transferred to the atoms resulting in ionisation and excitation, in contrast with the electrons we make the assumption that on the timescale of typical FEL pulses no kinetic energy is given to the ions, such that they are assumed to remain at room temperature throughout the calculation.  We consider such an assumption to be valid, given calculations that indicate that the timescale for electron-ion equilibriation, in terms of their temperatures, is several picoseconds~\cite{Ng1995, Matthieu2011, White2014}. 


CCFLY treats the atomic physics in terms of superconfigurations which denote the number of electrons with a specific principal quantum number, i.e. the number of electrons within the K, L or M shells.  In solid Mg, which is metallic, the M shell electrons are already effectively ionized, giving the ground state superconfiguration  (KLM) of (280).  This is taken into account by the introduction of a continuum lowering  (ionisation potential depression - IPD) routine  within the code.  Various models for such IPD have been put forward, with studies of the K-shell emission spectra from Al and Mg indicating that a modified version of the Ecker-Kroll IPD model~\cite{Ecker1963} provides results consistent with experiments under these conditions, at least for the first few charge states~\cite{Ciricosta2012,Ciricosta2016a}, although the exact model used does not affect the results presented here in any significant way.

The X-ray laser radiation field used in the calculations is assumed to be constant in time and Gaussian in frequency, with a fractional bandwidth of 0.4\% and a photon energy of 2000 eV. The intensity of the X-ray pulse is 3$\times 10^{17}$ W cm$^{-2}$.  In contrast to the work of Rohringer~\cite{Rohringer2007,Rohringer2009} we make no attempt in these initial studies to model the effect of the rapid temporal modulations in FEL intensity caused by the SASE (self amplified spontaneous emission) spikes, as our overarching aim is the more modest goal of elucidating the principal differences between the low and high density cases.

The ion density for solid Mg is 4.3 $\times 10^{22}$ cm$^{-3}$.  For comparisons with previous work, we also perform calculations at a density 4 orders of magnitude lower, at  4.3 $\times 10^{18}$  cm$^{-3}$, which is both sufficiently small so as to make collisional effects negligible, and is similar to the values previously used for the Neon gas targets.  For ease of comparison with the solid-density case, we also assume that the starting superconfiguration of the low density case is (280), i.e. that it is doubly-ionized before the onset of the FEL pulse (one could envisage the production of such a low density target by, for example, optical laser ablation of an Mg foil). 

\begin{figure*}[t]
\centering
\includegraphics[width=\linewidth]{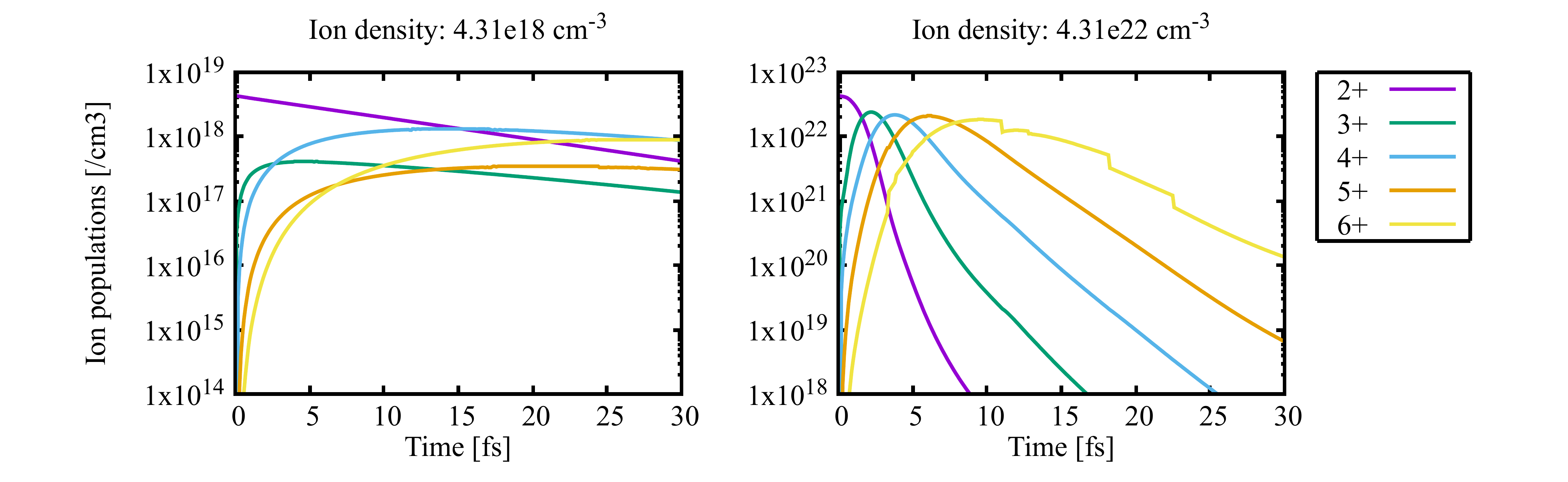}
\caption{The charge state distribution as a function of time for  low (left) and solid (right) density Mg.}\label{fig:Charge_States}
\end{figure*}

The simulations provide the populations of the various superconfigurations as a function of time, which in turn allow us to calculate the gain (or otherwise) of the relevant transitions.  The gain cross section for a particular transition, $\sigma_{stim}$, is given by
\begin{equation}
    \sigma_{stim} = \frac{2 \pi c^2 A}{\omega^2 \Delta \omega} \quad,
    \label{eq:sigma_stim}
\end{equation}
where the $A$ is the spontaneous emission rate in s$^{-1}$, and $\Delta \omega$ the width of the transition.  For the low density case this width will be dominated by the Auger lifetime (which, for example, for the transition from the (180) to (270) superconfigurations results in a width of order 0.4 eV).  However, at solid densities the line-width can increase due to collisional ionisation.  We take this into account in a very simple model where we assume an additional line width which is a function of the sum of the inverse of the relevant bound-bound and bound-free collisional rates for the upper and lower levels.  As we shall see below, the precise form of this additional width does not play a significant role in affecting the overall gain in these calculations.
The gain per atom is given by
\begin{equation}
    g(t) = \left( N_u(t) - N_l(t)*\frac{g_u}{g_l} \right )\frac{\sigma_{stim}(t) }{ N_i} \quad ,
    \label{eq:gain}
\end{equation}
where $N_u(t)$ and $N_l(t)$ are the populations of the upper and lower state at time $t$; $g_u$ and $g_l$ are the respective degeneracies, and $N_i$ is the total ion density.  Note the gain cross section is now considered to be time-dependent owing to the effect of collisions on the linewidth.

\section{Results}

Before turning our attention to lasing itself, in Fig. \ref{fig:Charge_States} we plot the populations of the ground state ions (i.e. those with 2 electrons in the K shell, and varying numbers of L shell electrons) as a function of time for both the low and high density cases.  It can immediately be seen that in the low density case the charge states that predominate have an even number of L shell electrons, whereas in the solid density case, as time proceeds, all of the successive charge states are produced.  For example, at a time of 30 fs in the low density case charge states 4+ and 6+ have populations several times greater than those of 3+ and 5+.  This alternation in the populations of the charge states in the low density case was demonstrated in one of the first experiments ever to be performed at LCLS, where the charge states produced when a Neon gas was irradiated by the FEL pulse were measured~\cite{Young2010}.  The results agreed well with simulations, including those undertaken using a previous version of the code used here~\cite{Ciricosta2011}.  This phenomenon is due to the fact that (assuming the FEL x-ray energy chosen lies above the K-shell photoionisation energy for all of the charge states to be considered),  after each K-shell photoionisation event, creating a K-shell hole, the dominant decay mechanism will be via the Auger effect, resulting in the filling of the K-shell hole by an L electron, and the ejection of another L electron into the continuum: thus the absorption of the FEL photon results in double ionisation.  In contrast, for the solid density case, the high electron density can cause collisional ionisation of the L-shell, resulting in all of the ground state ions being produced, as can readily be seen in Fig. \ref{fig:Charge_States}. This lack of alternation in the charges states produced in the solid density case was commented upon in work reporting the first observation of x-ray spectra from solid targets irradiated by the focussed output of the LCLS~\cite{Vinko2012}. Indeed, as we shall see, it is this very effect that reduces the gain per atom in the solid density case.  Photoionisation of the (280) atom creates the (180) superconfiguration, which is the upper state of the K-shell lasing transition which has a lower state of (270).  As the relative number of (270) ions is much greater in the solid density case, for the reasons given above, the creation of conditions conducive to lasing is much more difficult.

\begin{figure*} [t]
\centering
\includegraphics[width=\linewidth]{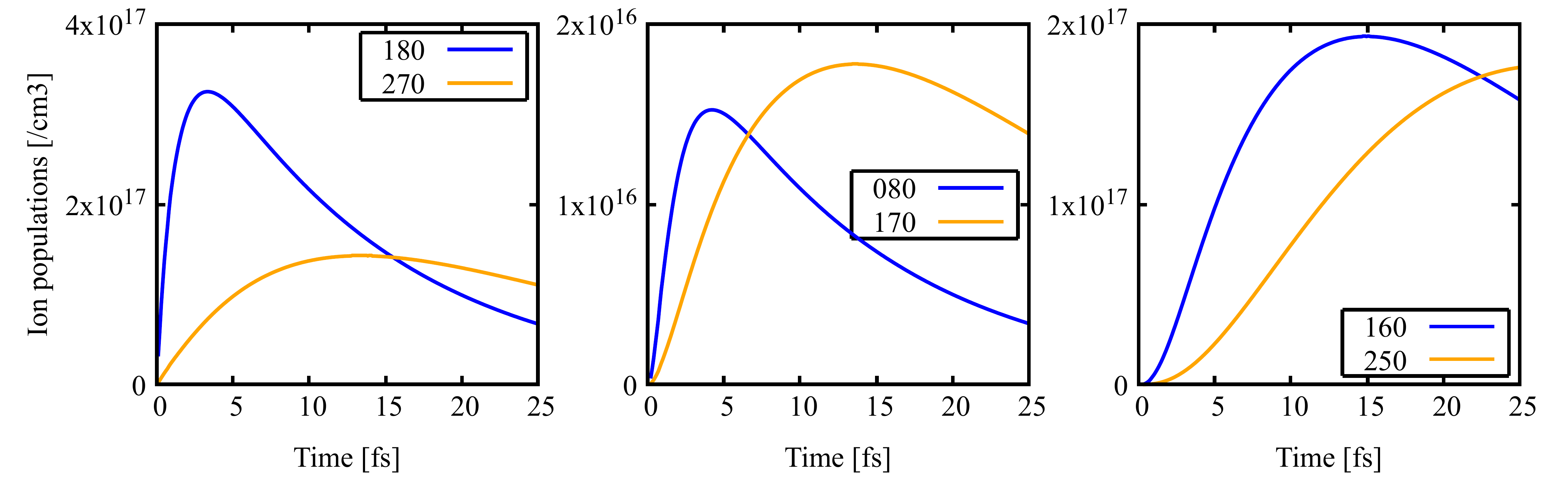}
\caption{Low density: Ion populations of the lower and upper states as a function of time}\label{fig:low_dens_ion_pop}
\end{figure*}

\begin{figure}
\centering
\includegraphics[width=\columnwidth]{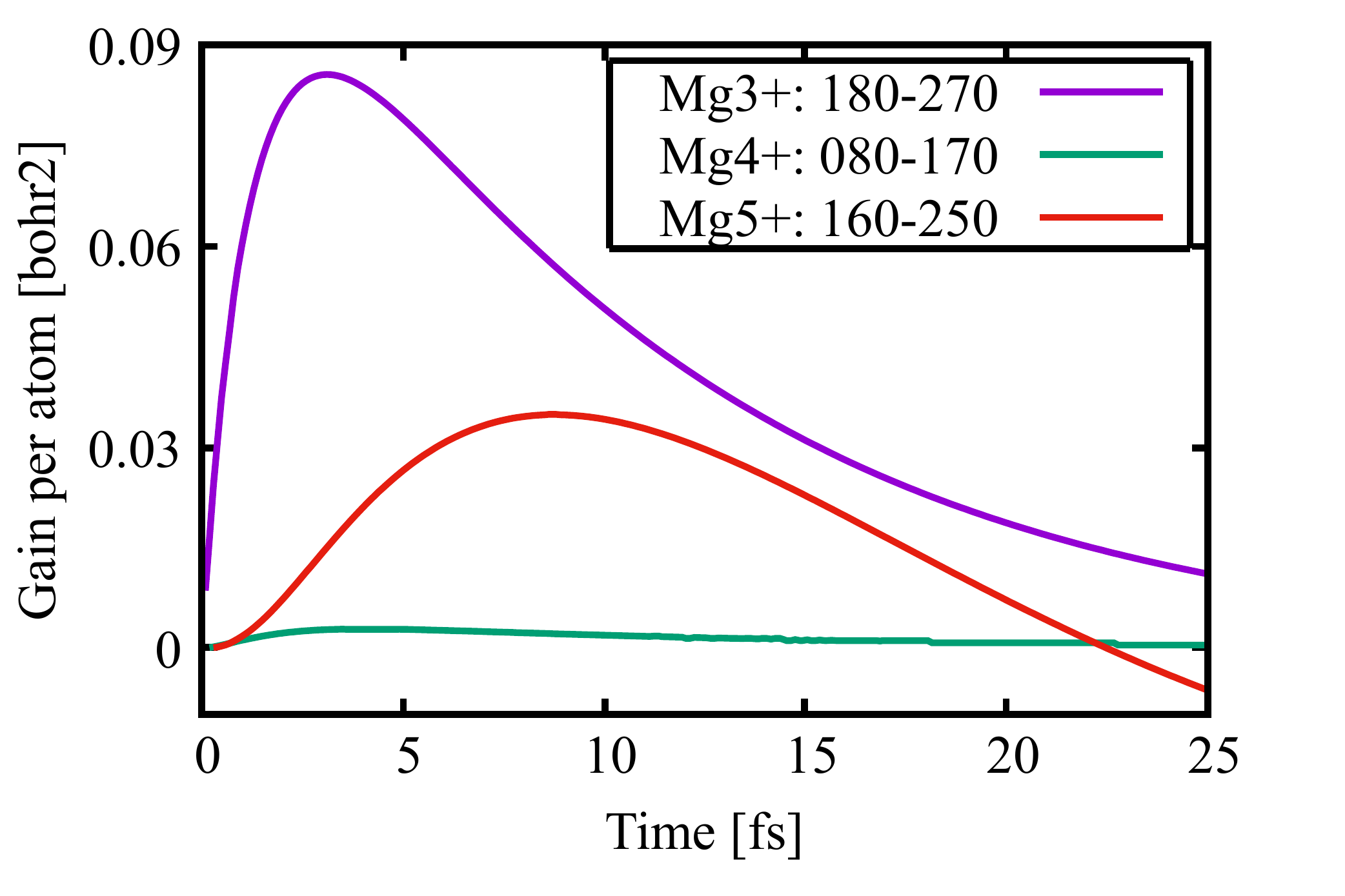}
\caption{Low density: Gain per atom of the transitions}\label{fig:Mg_low_dens_gain_per_atom}
\end{figure}

In order to illustrate this further we show In Fig. \ref{fig:low_dens_ion_pop} the populations as a function of time, in the low density case, of (180) and (270), (080) and (170) (given that it is possible to lase due to the creation of double core holes) and lastly, (160) and (250).  The corresponding gains per atom are shown in Fig. \ref{fig:Mg_low_dens_gain_per_atom}  in units of (bohr)$^2$. It can be seen that for Mg3+, lasing on the (180)-(270) transition, the gain per atom peaks after approximately 3-fsec at a value of 0.085 bohr$^2$.  This gain value is very similar to that found by Rohringer {\it et al.} in their numerical studies of lasing in Neon, and given the atomic numbers are close, and the pumping conditions not too dissimilar, such agreement is encouraging.  Furthermore, these values are also consistent with the gain values found in other simulations of the Neon system where at atomic densities of 2$\times 10^{19}$ cm$^{-3}$ gains of order 60 cm$^{-1}$ were predicted~\cite{Nilsen2016} (for this atomic density, our prediction for Mg would correspond to of order 40 cm$^{-1}$).  Note however that the gain values that we quote are for a superconfiguration, that is to say the gains for transitions between particular configurations have been summed (e.g. our gain for (180)-(270) is effectively the sum of gains for K$\alpha 1$ and K$\alpha 2$). 

We note that, as found by Rohringer, the system can also lase on the K$\alpha$ transitions of higher charge states.  Owing to the charge states with significant populations alternating due to Auger decay, as described above, the next K$\alpha$ transition with significant gain is that between the (160) and (250) superconfigurations, where the peak gain is slightly less than half that of the (180)-(270) transition.  We also note that owing to the high photoionisation rate of the K-shell, a small gain per atom is predicted on the transition with an upper state containing a double core hole: (080)-(170), but at a peak value of 0.0028 (bohr)$^2$, at these densities this would only translate to a gain of order 1.3 cm$^{-1}$, which. whilst it could prove difficult to experimentally verify, may still be measurable.

\begin{figure*}[t]
\centering
\includegraphics[width=\linewidth]{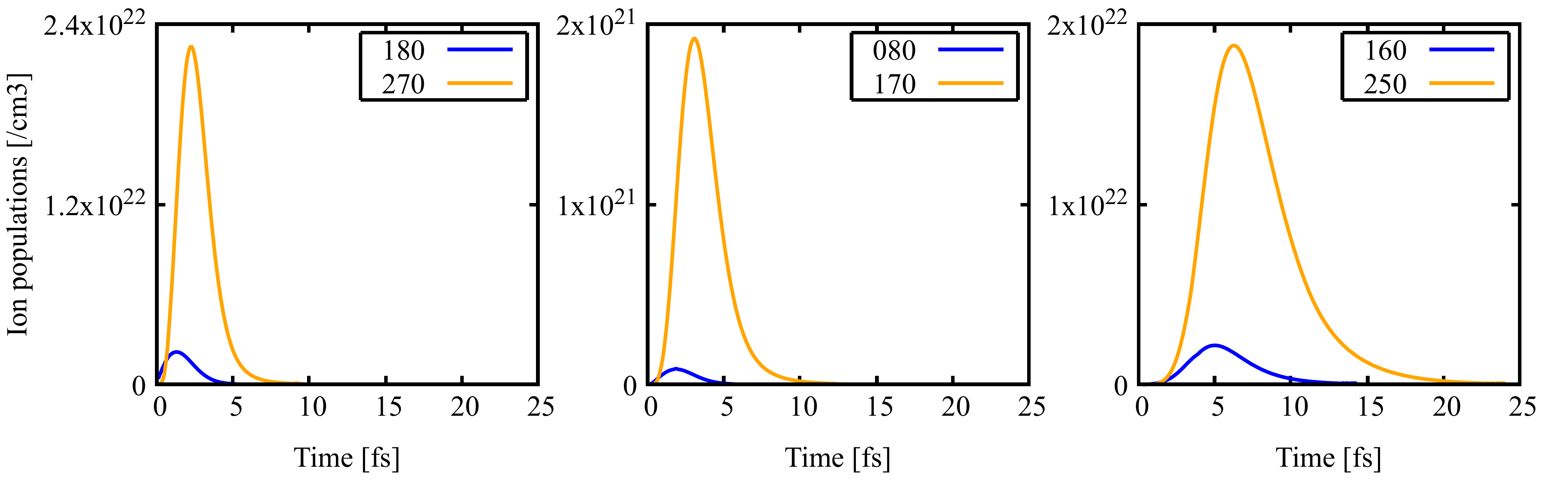}
\caption{Solid density: Ion populations of the lower and upper states as a function of time}\label{fig:high_dens_ion_pop}
\end{figure*}

\begin{figure}
\centering
\includegraphics[width=\columnwidth]{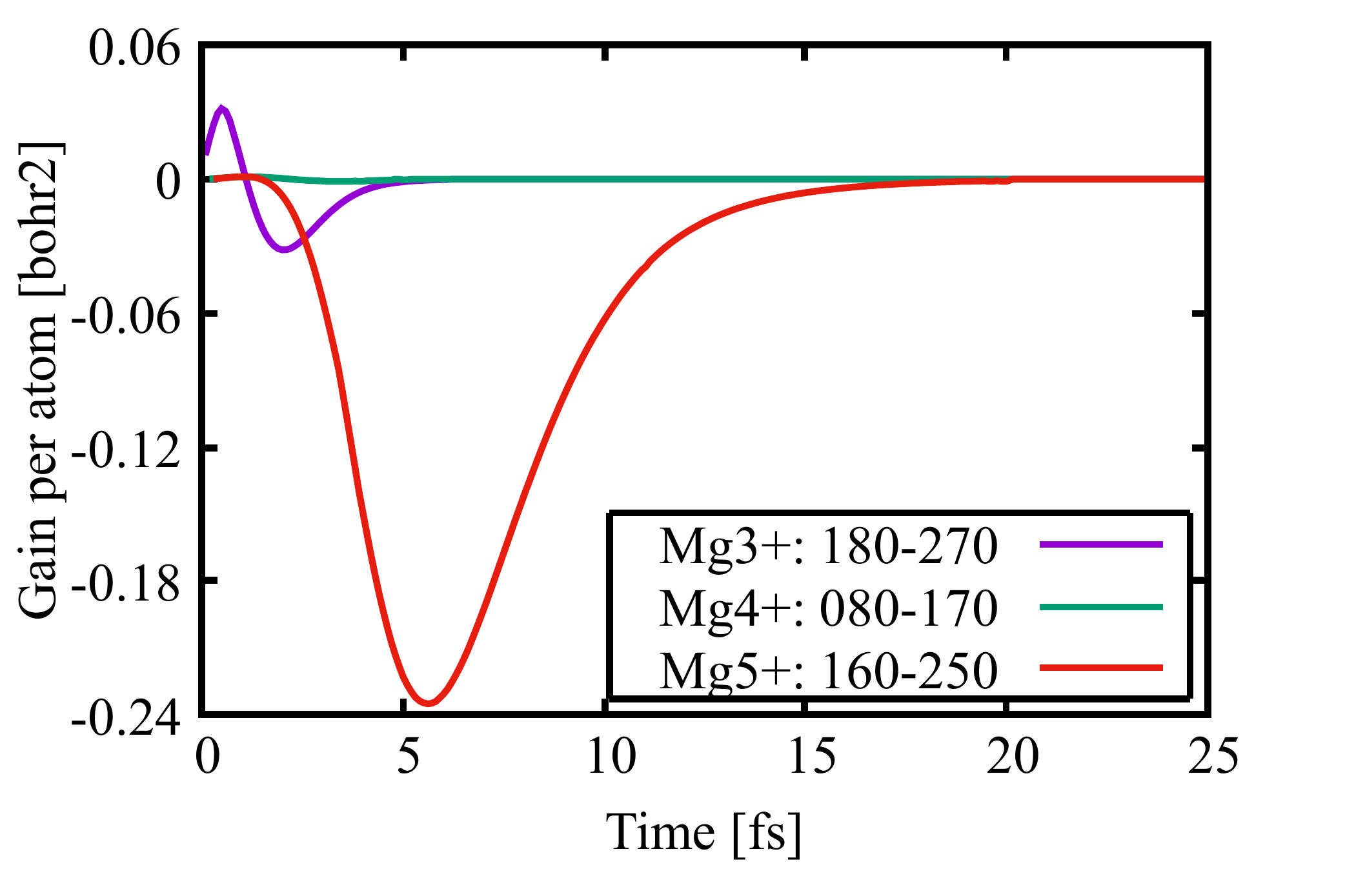}
\caption{Solid density: Gain per atom of the transitions}\label{fig:Mg_high_dens_gain_per_atom}
\end{figure}

This situation should be contrasted with that of the solid density case, where the populations of the relevant superconfigurations as a function of time are shown in Fig. \ref{fig:high_dens_ion_pop}.  Whilst here K-shell photoionisation starts to produce the upper state (180), as noted above the electrons that are ejected in this process can rapidly ionise the L-shell of (280), producing copious quantities of ions in the (270) superconfiguration, which is the lower state of the first lasing transition.  Thus we can see in Fig. \ref{fig:high_dens_ion_pop} that on a timescale of order a femtosecond the ground state population exceeds that of the upper state to such a degree that even taking account the different degeneracies the gain quickly goes negative, as shown in Fig. \ref{fig:Mg_high_dens_gain_per_atom}.  Indeed, the effect of collisions is such that for all of the subsequent charge states no gain whatsoever is observed.  For the first ion stage (i.e. the initial ion in the cold metal having had 1 K-shell electron photoionised) the gain per atom peaks at 0.024 bohr$^2$, more than a factor of 3 lower than in the low density case, and this is even with us assuming an instantaneous turning on of the FEL.  These results clearly demonstrate the more stringent requirements for lasing in the solid density case.  Indeed, the very short duration of the gain, even in the situation where within the simulation we have assumed instantaneous turn-on of the x-ray pulse, may provide some explanation as to why, for a given total energy in the x-ray pump, there are far greater variations in the gain seen experimentally in the liquid case than in the experiments with Neon gas: for example compare Fig. 3 of reference \cite{Rohringer2012} with Fig. 1(c) of reference \cite{Kroll2018}.

\begin{figure*}
\centering
\includegraphics[width=\linewidth]{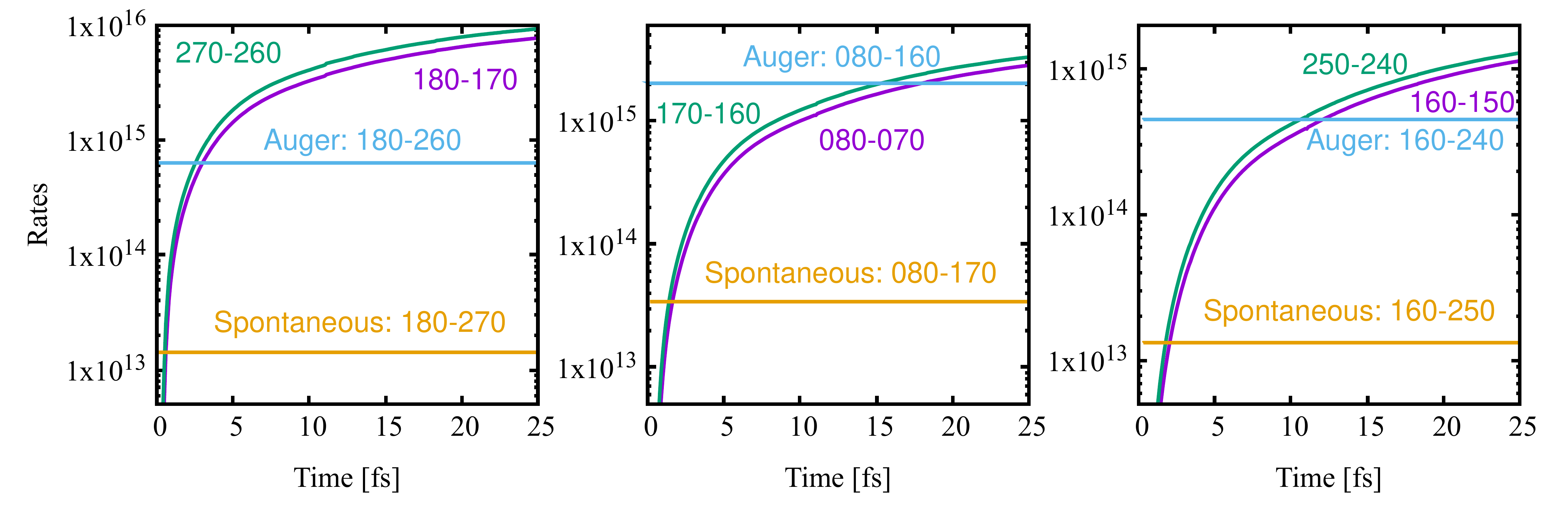}
\caption{Solid density: The spontaneous, Auger, and collisional rates as a function of time for transitions between various superconfigurations. }\label{fig:high_density_rates}
\end{figure*}

\begin{figure}
\centering
\includegraphics[width=\linewidth]{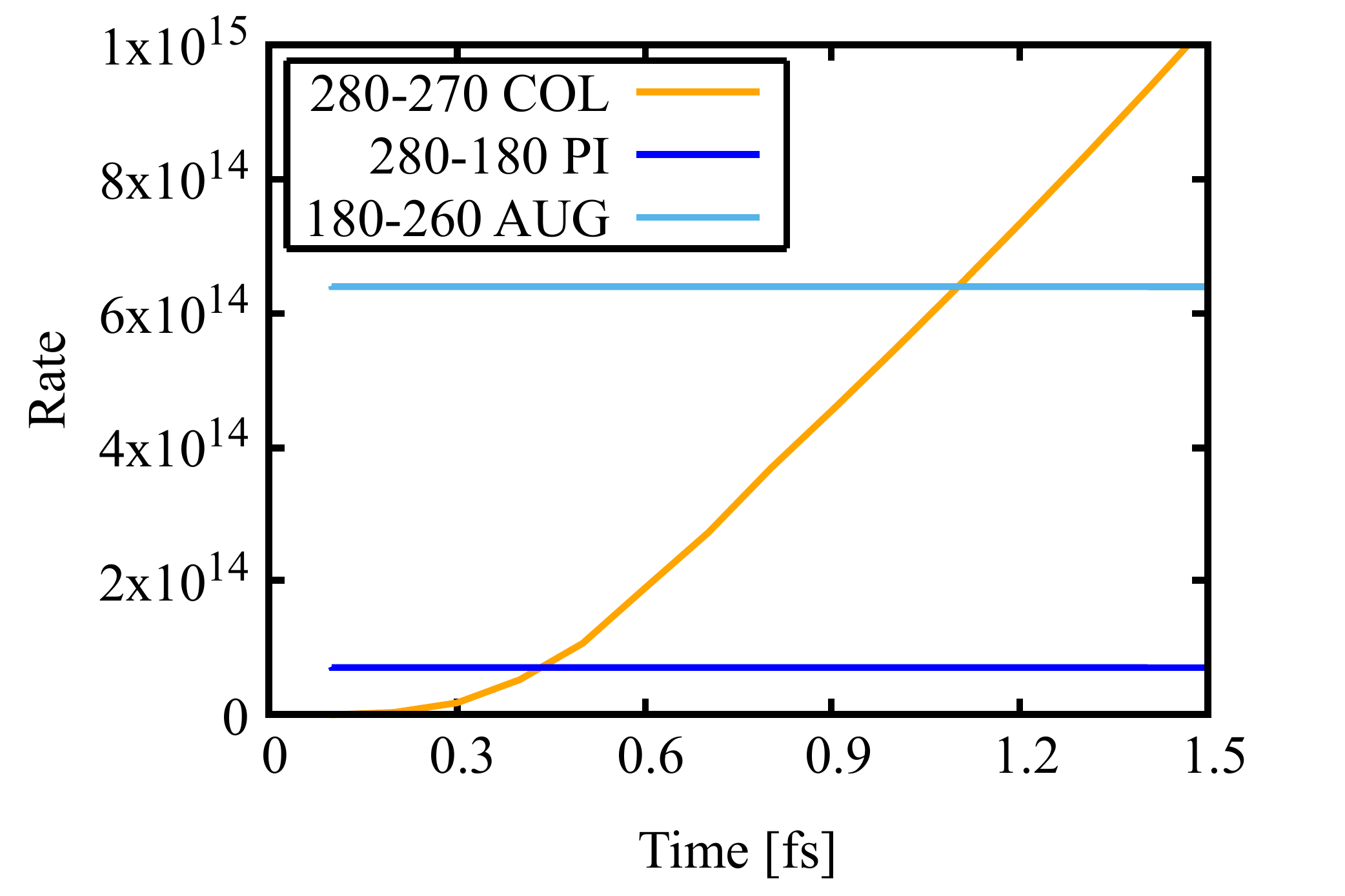}
\caption{Solid density: Collisional rate of 280-270, the photoionization rate of (280)-(170) due to the FEL and the Auger rate (180)-(260) as a function of time. }\label{fig:280_rates}
\end{figure}

Further insight into the above processes can be gleaned from plots showing the relevant rates as a function of time for the various transitions pertinent to the lasing processes; these are plotted in Fig. \ref{fig:high_density_rates} and Fig. \ref{fig:280_rates}.  From Fig. \ref{fig:high_density_rates} we see that the Auger rate would dominate in the absence of collisions, and is nearly two orders of magnitude greater than the spontaneous rate (with the ratio of the spontaneous to Auger rate being the fluorescence yield).  We note that for all three transitions shown the collisional ionisation rates rapidly dominate over the Auger rate.  In particular, and as shown in Fig. \ref{fig:280_rates} we see that the collisional ionisation rate that produces the lower state of the first lasing transition, i.e the superconfiguration (270), exceeds the Auger rate of (180)-(260) on a timescale of order a femtosecond.

Thus far we have considered the curtailing of the gain in the solid density case as being due to collisional effects rapidly populating the ground state of the transition.  However, as noted above, such collisions also reduced the gain cross section by increasing the line-width of the transition.  The relative importance of these two effects can be seen in Fig. \ref{fig:compare}, where we show the gain per atom on the (180)-(270) transition, the linewidth, and the effective population inversion (i.e. $[ N_u(t) - N_l(t)*g_u/g_l]/N_i]$) as a function of time.  It can be seen that the peak gain occurs very slightly earlier (about 0.1 fs) than the peak in the inversion ratio, and that this is due to the increase in the line-width.  However, the curtailment of the gain is clearly dominated by the changes in the populations, that is to say the inversion ratio itself, with the increasing width of the lasing transition only playing a minor role.

A further difference, however, between the low and high density cases is the duration of positive gain: 11 fs (FWHM) in the low density case for the (180)-(270) transition, yet just 0.65 fs (FWHM) in the solid density case.  We see that whilst the peak gain per atom is reduced owing to the collisional effects, this very same mechanism also results in a reduced pulse length of the laser, which may well have practical advantages if such systems are subsequently to be used in further applications, especially as the pulse length appears to be approaching that of the standard unit of atomic time, $a_0/(\alpha c)$ = 24.3 as.

\section{Discussion}

There are several directions which appear fruitful for further study.  Given the extremely short timescales over which gain in the solid state system exists owing to collisional effects it is evident that any improvements in understanding of the relevant rates could well influence the degree and duration of gain predicted.  It should be noted that the rates used in this study have been the those used in previous versions of  the SCFLY code, whereas recent  experimental work has shown that these rates may well be an underestimate~\cite{Vinko2015,Vandenberg2018}.  Furthermore, here the assumption has also been made that the distribution function of the electrons in the continuum is at all times Maxwellian - i.e. we have both ignored the fact that the cold metal will have a Fermi-Dirac distribution of the electrons, and we have further assumed  that thermalisation of the electrons is instantaneous.  

\begin{figure}
\centering
\includegraphics[width=\linewidth]{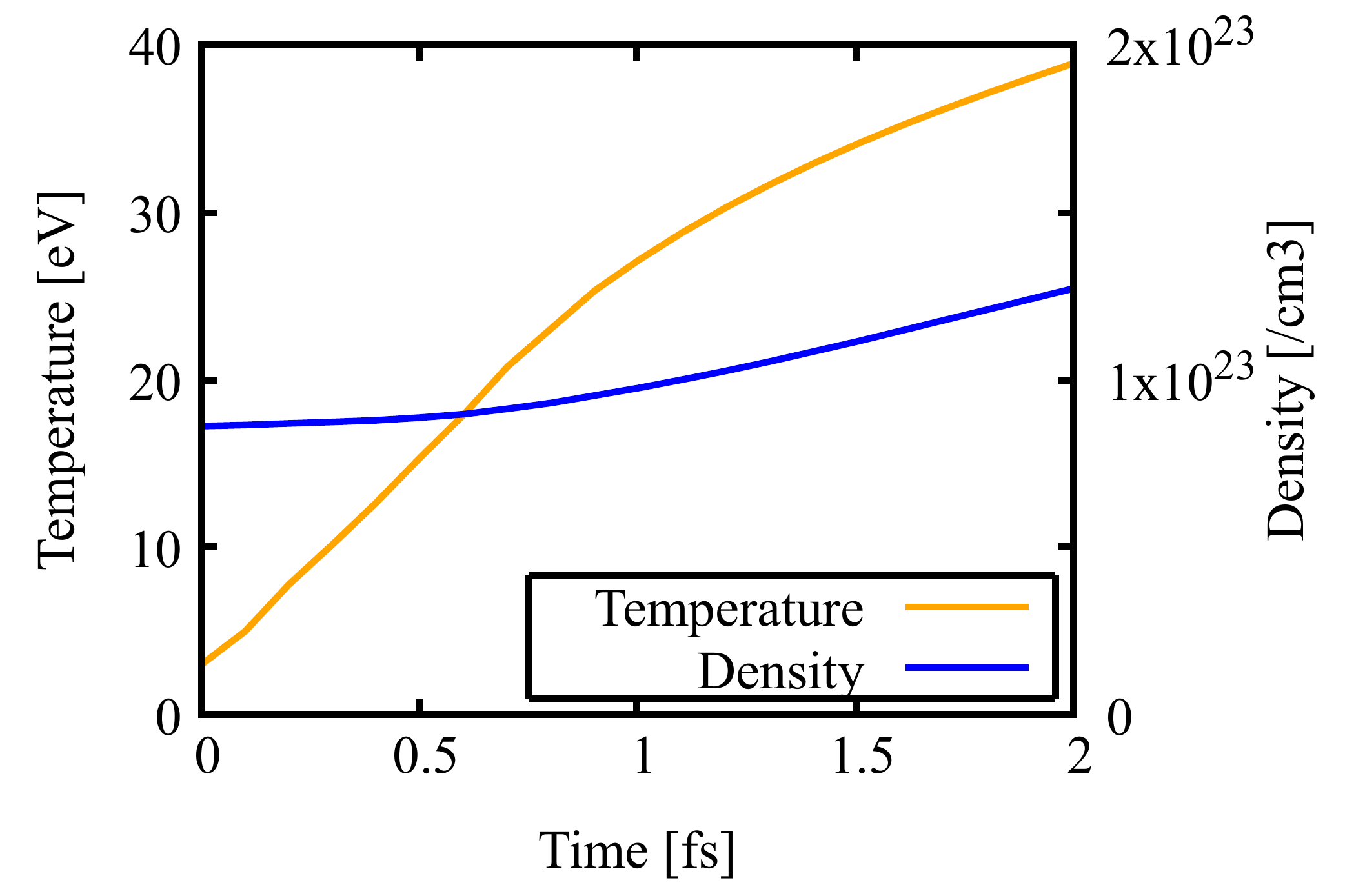}
\caption{Solid density: Temperature and electron number density as a function of time. }\label{fig:Temp_Dens}
\end{figure}

\begin{figure}
\centering
\includegraphics[width=\linewidth]{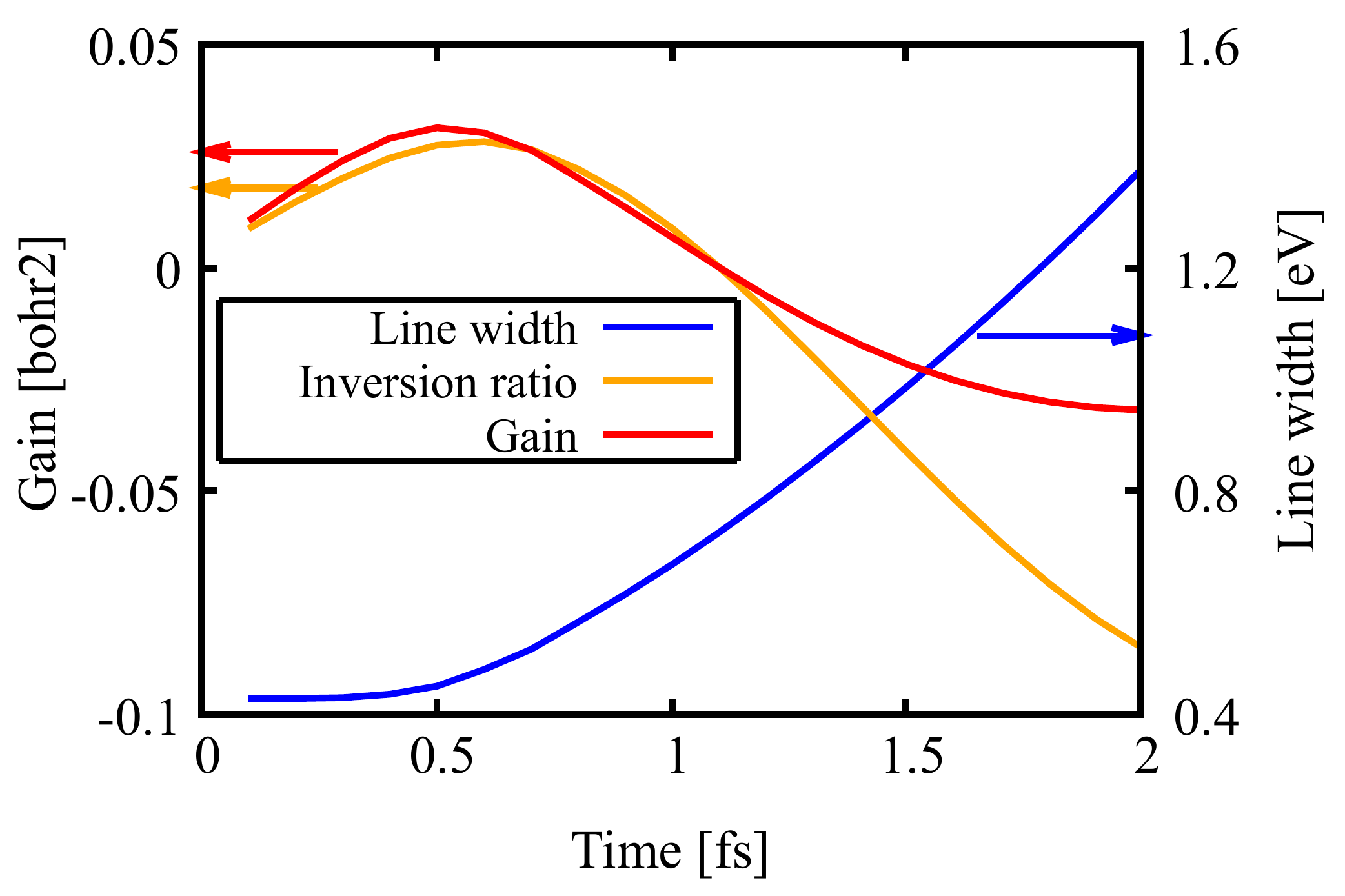}
\caption{Solid density: the gain (left axis in units of bohr$^2$, inversion ratio (left axis in dimensionless units) and line width (right axis in units of eV) as a function of time for the (180)-(270) transition.}\label{fig:compare}
\end{figure}

Both of the above assumptions warrant scrutiny.  If, for the sake of argument, we do assume instantaneous thermalisation, the fact that the current version of the code does not take into account the Fermi degeneracy of the initial electrons is unlikely to be significant because  the electron temperature exceeds the Fermi temperature on exceedingly short timescales.  Fig. \ref{fig:Temp_Dens} shows the calculated electron temperature and density as a function of time in the solid density case. The electron temperature rise exceeds the Fermi energy of Mg (4.4 eV), justifying classical statistics in less than 0.2 fs.

However, such is the brevity of the gain in the solid system that it may well be that the assumption of instantaneous thermalisation of the electrons needs to be reconsidered, given that electrons are ejected into the continuum via photoionization and Auger decay at specific energies.  A more recent version of the code used here has been developed that, with considerable increased computational cost, keeps track of the evolving non-thermal electron distribution function, and alters the collisional rates used accordingly~\cite{Shenyuan2022}. Further studies incorporating this capability would clearly be beneficial.  Extending studies to materials of higher atomic number (again at potentially greater computation cost) would also be of interest.  We also note that owing to the extremely short duration of the predicted gain in the solid density case, we might very well expect considerable differences in gain if we relax the assumption of a top-hat FEL pulse - that is to say the statistics of the SASE spikes, considered by Rohringer in the low density case~\cite{Rohringer2007,Rohringer2009}, might well play an equal if not even more important role in the statistics of the K$\alpha$ lasing output in the solid density case.  Finally, the current version of CCFLY treats the system in terms of superconfigurations.  Whilst tracking the populations of the myriad of possible configurations as a function of time would be computationally prohibitive in many simulations of FEL-matter interactions, in the case of K-shell lasing in solid or liquid density systems, the work presented here indicates that only the first one or two charge states will be of any real significance.  Thus a more detailed treatment of the atomic physics, including collisional effects, of these few states may well be feasible in future studies.



\section{Acknowledgements}

JSW and SMV acknowledge support from UK EPSRC grants EP/P015794/1 and EP/W010097/1. JSW is grateful for further support from OxCHEDS (the Oxford Centre for High Energy Density Science). SMV is a Royal Society University Research Fellow.

\section{Appendix}
From the results presented in this paper it is clear that electron collisional ionization plays an important role in determining the lifetime of states that may show gain. We expand on the details of our modelling approach here.

The electron collisional ionization rate $S$ of some ion is given by the average of the collisional ionization cross section $\sigma$ over the distribution of available electron velocities $v$, i.e., $S = \langle \sigma v \rangle$. In our modelling we assume a Maxwellian distribution for the free electron population, and that this distribution thermalizes instantaneously at each time step of the collisional-radiative calculation. Then, we can write the rate explicitly as
\begin{equation}
    S = n_e \sqrt{\frac{8}{\pi m_e}} (k_B T)^{-3/2} \int_0^{\infty} \varepsilon \, \sigma(\varepsilon)\, e^{-\varepsilon / k_B T} {\rm d}\varepsilon.
\end{equation}
Here $n_e$ is the density, $T$ the temperature, and $m_e$ the mass of the free electrons. The integral is taken over the electron energy. There are many possible choices for the collisional ionization cross section for dense systems (for example, see Ref.~\cite{vandenberg:2018} and references therein for a more detailed discussion). In this work we have used the simple empirical Lotz model~\cite{1967ZPhy..206..205L}, where the energy-dependent collisional cross section is given by the expression
\begin{equation}
    \sigma (\varepsilon) \approx \sum_i a_i q_i \frac{\ln(\varepsilon/P_i)}{\varepsilon P_i},
\end{equation}
where the sum is over all atomic sub-shells. The $a_i$ terms are fitting constants~\cite{lotz:1968}, $q_i$ denotes the population of the sub-shell, and $P_i$ is the ionization energy of a particular sub-shell. The effect of continuum lowering can be included in the model by an appropriate re-scaling of the ionization energy.
We have verified that the cross sections calculated from this simple model are comparable to more advanced calculations based on the Binary Encounter, Distorted Wave, or Scaled Hydrogenic models~\cite{vandenberg:2018}.

\bibliography{jswBIB-2a}

\begin{thebibliography}{35}%
\makeatletter
\providecommand \@ifxundefined [1]{%
 \@ifx{#1\undefined}
}%
\providecommand \@ifnum [1]{%
 \ifnum #1\expandafter \@firstoftwo
 \else \expandafter \@secondoftwo
 \fi
}%
\providecommand \@ifx [1]{%
 \ifx #1\expandafter \@firstoftwo
 \else \expandafter \@secondoftwo
 \fi
}%
\providecommand \natexlab [1]{#1}%
\providecommand \enquote  [1]{``#1''}%
\providecommand \bibnamefont  [1]{#1}%
\providecommand \bibfnamefont [1]{#1}%
\providecommand \citenamefont [1]{#1}%
\providecommand \href@noop [0]{\@secondoftwo}%
\providecommand \href [0]{\begingroup \@sanitize@url \@href}%
\providecommand \@href[1]{\@@startlink{#1}\@@href}%
\providecommand \@@href[1]{\endgroup#1\@@endlink}%
\providecommand \@sanitize@url [0]{\catcode `\\12\catcode `\$12\catcode
  `\&12\catcode `\#12\catcode `\^12\catcode `\_12\catcode `\%12\relax}%
\providecommand \@@startlink[1]{}%
\providecommand \@@endlink[0]{}%
\providecommand \url  [0]{\begingroup\@sanitize@url \@url }%
\providecommand \@url [1]{\endgroup\@href {#1}{\urlprefix }}%
\providecommand \urlprefix  [0]{URL }%
\providecommand \Eprint [0]{\href }%
\providecommand \doibase [0]{https://doi.org/}%
\providecommand \selectlanguage [0]{\@gobble}%
\providecommand \bibinfo  [0]{\@secondoftwo}%
\providecommand \bibfield  [0]{\@secondoftwo}%
\providecommand \translation [1]{[#1]}%
\providecommand \BibitemOpen [0]{}%
\providecommand \bibitemStop [0]{}%
\providecommand \bibitemNoStop [0]{.\EOS\space}%
\providecommand \EOS [0]{\spacefactor3000\relax}%
\providecommand \BibitemShut  [1]{\csname bibitem#1\endcsname}%
\let\auto@bib@innerbib\@empty
\bibitem [{\citenamefont {Duguay}\ and\ \citenamefont
  {Rentzepis}(1967)}]{Duguay1967}%
  \BibitemOpen
  \bibfield  {author} {\bibinfo {author} {\bibfnamefont {M.~A.}\ \bibnamefont
  {Duguay}}\ and\ \bibinfo {author} {\bibfnamefont {P.~M.}\ \bibnamefont
  {Rentzepis}},\ }\bibfield  {title} {\enquote {\bibinfo {title} {Some
  approaches to vacuum uv and x-ray lasers},}\ }\href
  {https://doi.org/10.1063/1.1728208} {\bibfield  {journal} {\bibinfo
  {journal} {Applied Physics Letters}\ }\textbf {\bibinfo {volume} {10}},\
  \bibinfo {pages} {350--352} (\bibinfo {year} {1967})},\ \Eprint
  {https://arxiv.org/abs/https://doi.org/10.1063/1.1728208}
  {https://doi.org/10.1063/1.1728208} \BibitemShut {NoStop}%
\bibitem [{\citenamefont {Rohringer}\ \emph {et~al.}(2012)\citenamefont
  {Rohringer}, \citenamefont {Ryan}, \citenamefont {London}, \citenamefont
  {Purvis}, \citenamefont {Albert}, \citenamefont {Dunn}, \citenamefont
  {Bozek}, \citenamefont {Bostedt}, \citenamefont {Graf}, \citenamefont {Hill},
  \citenamefont {Hau-Riege},\ and\ \citenamefont {Rocca}}]{Rohringer2012}%
  \BibitemOpen
  \bibfield  {author} {\bibinfo {author} {\bibfnamefont {N.}~\bibnamefont
  {Rohringer}}, \bibinfo {author} {\bibfnamefont {D.}~\bibnamefont {Ryan}},
  \bibinfo {author} {\bibfnamefont {R.~A.}\ \bibnamefont {London}}, \bibinfo
  {author} {\bibfnamefont {M.}~\bibnamefont {Purvis}}, \bibinfo {author}
  {\bibfnamefont {F.}~\bibnamefont {Albert}}, \bibinfo {author} {\bibfnamefont
  {J.}~\bibnamefont {Dunn}}, \bibinfo {author} {\bibfnamefont {J.~D.}\
  \bibnamefont {Bozek}}, \bibinfo {author} {\bibfnamefont {C.}~\bibnamefont
  {Bostedt}}, \bibinfo {author} {\bibfnamefont {A.}~\bibnamefont {Graf}},
  \bibinfo {author} {\bibfnamefont {R.}~\bibnamefont {Hill}}, \bibinfo {author}
  {\bibfnamefont {S.~P.}\ \bibnamefont {Hau-Riege}},\ and\ \bibinfo {author}
  {\bibfnamefont {J.~J.}\ \bibnamefont {Rocca}},\ }\bibfield  {title} {\enquote
  {\bibinfo {title} {Atomic inner-shell x-ray laser at 1.46 nanometres pumped
  by an x-ray free-electron laser},}\ }\href
  {https://doi.org/10.1038/nature10721} {\bibfield  {journal} {\bibinfo
  {journal} {Nature}\ }\textbf {\bibinfo {volume} {481}},\ \bibinfo {pages}
  {488--491} (\bibinfo {year} {2012})}\BibitemShut {NoStop}%
\bibitem [{\citenamefont {Yoneda}\ \emph {et~al.}(2015)\citenamefont {Yoneda},
  \citenamefont {Inubushi}, \citenamefont {Nagamine}, \citenamefont {Michine},
  \citenamefont {Ohashi}, \citenamefont {Yumoto}, \citenamefont {Yamauchi},
  \citenamefont {Mimura}, \citenamefont {Kitamura}, \citenamefont {Katayama},
  \citenamefont {Ishikawa},\ and\ \citenamefont {Yabashi}}]{Yoneda2015}%
  \BibitemOpen
  \bibfield  {author} {\bibinfo {author} {\bibfnamefont {H.}~\bibnamefont
  {Yoneda}}, \bibinfo {author} {\bibfnamefont {Y.}~\bibnamefont {Inubushi}},
  \bibinfo {author} {\bibfnamefont {K.}~\bibnamefont {Nagamine}}, \bibinfo
  {author} {\bibfnamefont {Y.}~\bibnamefont {Michine}}, \bibinfo {author}
  {\bibfnamefont {H.}~\bibnamefont {Ohashi}}, \bibinfo {author} {\bibfnamefont
  {H.}~\bibnamefont {Yumoto}}, \bibinfo {author} {\bibfnamefont
  {K.}~\bibnamefont {Yamauchi}}, \bibinfo {author} {\bibfnamefont
  {H.}~\bibnamefont {Mimura}}, \bibinfo {author} {\bibfnamefont
  {H.}~\bibnamefont {Kitamura}}, \bibinfo {author} {\bibfnamefont
  {T.}~\bibnamefont {Katayama}}, \bibinfo {author} {\bibfnamefont
  {T.}~\bibnamefont {Ishikawa}},\ and\ \bibinfo {author} {\bibfnamefont
  {M.}~\bibnamefont {Yabashi}},\ }\bibfield  {title} {\enquote {\bibinfo
  {title} {Atomic inner-shell laser at 1.5-{\aa}ngstr{\"o}m wavelength pumped
  by an x-ray free-electron laser},}\ }\href
  {https://doi.org/10.1038/nature14894} {\bibfield  {journal} {\bibinfo
  {journal} {Nature}\ }\textbf {\bibinfo {volume} {524}},\ \bibinfo {pages}
  {446--449} (\bibinfo {year} {2015})}\BibitemShut {NoStop}%
\bibitem [{\citenamefont {Kroll}\ \emph {et~al.}(2018)\citenamefont {Kroll},
  \citenamefont {Weninger}, \citenamefont {Alonso-Mori}, \citenamefont
  {Sokaras}, \citenamefont {Zhu}, \citenamefont {Mercadier}, \citenamefont
  {Majety}, \citenamefont {Marinelli}, \citenamefont {Lutman}, \citenamefont
  {Guetg}, \citenamefont {Decker}, \citenamefont {Boutet}, \citenamefont
  {Aquila}, \citenamefont {Koglin}, \citenamefont {Koralek}, \citenamefont
  {DePonte}, \citenamefont {Kern}, \citenamefont {Fuller}, \citenamefont
  {Pastor}, \citenamefont {Fransson}, \citenamefont {Zhang}, \citenamefont
  {Yano}, \citenamefont {Yachandra}, \citenamefont {Rohringer},\ and\
  \citenamefont {Bergmann}}]{Kroll2018}%
  \BibitemOpen
  \bibfield  {author} {\bibinfo {author} {\bibfnamefont {T.}~\bibnamefont
  {Kroll}}, \bibinfo {author} {\bibfnamefont {C.}~\bibnamefont {Weninger}},
  \bibinfo {author} {\bibfnamefont {R.}~\bibnamefont {Alonso-Mori}}, \bibinfo
  {author} {\bibfnamefont {D.}~\bibnamefont {Sokaras}}, \bibinfo {author}
  {\bibfnamefont {D.}~\bibnamefont {Zhu}}, \bibinfo {author} {\bibfnamefont
  {L.}~\bibnamefont {Mercadier}}, \bibinfo {author} {\bibfnamefont {V.~P.}\
  \bibnamefont {Majety}}, \bibinfo {author} {\bibfnamefont {A.}~\bibnamefont
  {Marinelli}}, \bibinfo {author} {\bibfnamefont {A.}~\bibnamefont {Lutman}},
  \bibinfo {author} {\bibfnamefont {M.~W.}\ \bibnamefont {Guetg}}, \bibinfo
  {author} {\bibfnamefont {F.-J.}\ \bibnamefont {Decker}}, \bibinfo {author}
  {\bibfnamefont {S.}~\bibnamefont {Boutet}}, \bibinfo {author} {\bibfnamefont
  {A.}~\bibnamefont {Aquila}}, \bibinfo {author} {\bibfnamefont
  {J.}~\bibnamefont {Koglin}}, \bibinfo {author} {\bibfnamefont
  {J.}~\bibnamefont {Koralek}}, \bibinfo {author} {\bibfnamefont {D.~P.}\
  \bibnamefont {DePonte}}, \bibinfo {author} {\bibfnamefont {J.}~\bibnamefont
  {Kern}}, \bibinfo {author} {\bibfnamefont {F.~D.}\ \bibnamefont {Fuller}},
  \bibinfo {author} {\bibfnamefont {E.}~\bibnamefont {Pastor}}, \bibinfo
  {author} {\bibfnamefont {T.}~\bibnamefont {Fransson}}, \bibinfo {author}
  {\bibfnamefont {Y.}~\bibnamefont {Zhang}}, \bibinfo {author} {\bibfnamefont
  {J.}~\bibnamefont {Yano}}, \bibinfo {author} {\bibfnamefont {V.~K.}\
  \bibnamefont {Yachandra}}, \bibinfo {author} {\bibfnamefont {N.}~\bibnamefont
  {Rohringer}},\ and\ \bibinfo {author} {\bibfnamefont {U.}~\bibnamefont
  {Bergmann}},\ }\bibfield  {title} {\enquote {\bibinfo {title} {Stimulated
  x-ray emission spectroscopy in transition metal complexes},}\ }\href
  {https://doi.org/10.1103/PhysRevLett.120.133203} {\bibfield  {journal}
  {\bibinfo  {journal} {Phys. Rev. Lett.}\ }\textbf {\bibinfo {volume} {120}},\
  \bibinfo {pages} {133203} (\bibinfo {year} {2018})}\BibitemShut {NoStop}%
\bibitem [{\citenamefont {Matthews}\ \emph {et~al.}(1985)\citenamefont
  {Matthews}, \citenamefont {Hagelstein}, \citenamefont {Rosen}, \citenamefont
  {Eckart}, \citenamefont {Ceglio}, \citenamefont {Hazi}, \citenamefont
  {Medecki}, \citenamefont {MacGowan}, \citenamefont {Trebes}, \citenamefont
  {Whitten}, \citenamefont {Campbell}, \citenamefont {Hatcher}, \citenamefont
  {Hawryluk}, \citenamefont {Kauffman}, \citenamefont {Pleasance},
  \citenamefont {Rambach}, \citenamefont {Scofield}, \citenamefont {Stone},\
  and\ \citenamefont {Weaver}}]{Matthews1985}%
  \BibitemOpen
  \bibfield  {author} {\bibinfo {author} {\bibfnamefont {D.~L.}\ \bibnamefont
  {Matthews}}, \bibinfo {author} {\bibfnamefont {P.~L.}\ \bibnamefont
  {Hagelstein}}, \bibinfo {author} {\bibfnamefont {M.~D.}\ \bibnamefont
  {Rosen}}, \bibinfo {author} {\bibfnamefont {M.~J.}\ \bibnamefont {Eckart}},
  \bibinfo {author} {\bibfnamefont {N.~M.}\ \bibnamefont {Ceglio}}, \bibinfo
  {author} {\bibfnamefont {A.~U.}\ \bibnamefont {Hazi}}, \bibinfo {author}
  {\bibfnamefont {H.}~\bibnamefont {Medecki}}, \bibinfo {author} {\bibfnamefont
  {B.~J.}\ \bibnamefont {MacGowan}}, \bibinfo {author} {\bibfnamefont {J.~E.}\
  \bibnamefont {Trebes}}, \bibinfo {author} {\bibfnamefont {B.~L.}\
  \bibnamefont {Whitten}}, \bibinfo {author} {\bibfnamefont {E.~M.}\
  \bibnamefont {Campbell}}, \bibinfo {author} {\bibfnamefont {C.~W.}\
  \bibnamefont {Hatcher}}, \bibinfo {author} {\bibfnamefont {A.~M.}\
  \bibnamefont {Hawryluk}}, \bibinfo {author} {\bibfnamefont {R.~L.}\
  \bibnamefont {Kauffman}}, \bibinfo {author} {\bibfnamefont {L.~D.}\
  \bibnamefont {Pleasance}}, \bibinfo {author} {\bibfnamefont {G.}~\bibnamefont
  {Rambach}}, \bibinfo {author} {\bibfnamefont {J.~H.}\ \bibnamefont
  {Scofield}}, \bibinfo {author} {\bibfnamefont {G.}~\bibnamefont {Stone}},\
  and\ \bibinfo {author} {\bibfnamefont {T.~A.}\ \bibnamefont {Weaver}},\
  }\bibfield  {title} {\enquote {\bibinfo {title} {Demonstration of a soft
  x-ray amplifier},}\ }\href {https://doi.org/10.1103/PhysRevLett.54.110}
  {\bibfield  {journal} {\bibinfo  {journal} {Phys. Rev. Lett.}\ }\textbf
  {\bibinfo {volume} {54}},\ \bibinfo {pages} {110--113} (\bibinfo {year}
  {1985})}\BibitemShut {NoStop}%
\bibitem [{\citenamefont {Chenais-Popovics}\ \emph {et~al.}(1987)\citenamefont
  {Chenais-Popovics}, \citenamefont {Corbett}, \citenamefont {Hooker},
  \citenamefont {Key}, \citenamefont {Kiehn}, \citenamefont {Lewis},
  \citenamefont {Pert}, \citenamefont {Regan}, \citenamefont {Rose},
  \citenamefont {Sadaat}, \citenamefont {Smith}, \citenamefont {Tomie},\ and\
  \citenamefont {Willi}}]{Popovics1987}%
  \BibitemOpen
  \bibfield  {author} {\bibinfo {author} {\bibfnamefont {C.}~\bibnamefont
  {Chenais-Popovics}}, \bibinfo {author} {\bibfnamefont {R.}~\bibnamefont
  {Corbett}}, \bibinfo {author} {\bibfnamefont {C.~J.}\ \bibnamefont {Hooker}},
  \bibinfo {author} {\bibfnamefont {M.~H.}\ \bibnamefont {Key}}, \bibinfo
  {author} {\bibfnamefont {G.~P.}\ \bibnamefont {Kiehn}}, \bibinfo {author}
  {\bibfnamefont {C.~L.~S.}\ \bibnamefont {Lewis}}, \bibinfo {author}
  {\bibfnamefont {G.~J.}\ \bibnamefont {Pert}}, \bibinfo {author}
  {\bibfnamefont {C.}~\bibnamefont {Regan}}, \bibinfo {author} {\bibfnamefont
  {S.~J.}\ \bibnamefont {Rose}}, \bibinfo {author} {\bibfnamefont
  {S.}~\bibnamefont {Sadaat}}, \bibinfo {author} {\bibfnamefont
  {R.}~\bibnamefont {Smith}}, \bibinfo {author} {\bibfnamefont
  {T.}~\bibnamefont {Tomie}},\ and\ \bibinfo {author} {\bibfnamefont
  {O.}~\bibnamefont {Willi}},\ }\bibfield  {title} {\enquote {\bibinfo {title}
  {Laser amplification at 18.2 nm in recombining plasma from a laser-irradiated
  carbon fiber},}\ }\href {https://doi.org/10.1103/PhysRevLett.59.2161}
  {\bibfield  {journal} {\bibinfo  {journal} {Phys. Rev. Lett.}\ }\textbf
  {\bibinfo {volume} {59}},\ \bibinfo {pages} {2161--2164} (\bibinfo {year}
  {1987})}\BibitemShut {NoStop}%
\bibitem [{\citenamefont {Matthews}(1995)}]{Matthews1995}%
  \BibitemOpen
  \bibfield  {author} {\bibinfo {author} {\bibfnamefont {D.~L.}\ \bibnamefont
  {Matthews}},\ }\bibfield  {title} {\enquote {\bibinfo {title} {Review of
  x-ray lasers},}\ }\href
  {https://doi.org/https://doi.org/10.1016/0168-583X(95)00079-8} {\bibfield
  {journal} {\bibinfo  {journal} {Nuclear Instruments and Methods in Physics
  Research Section B: Beam Interactions with Materials and Atoms}\ }\textbf
  {\bibinfo {volume} {98}},\ \bibinfo {pages} {91--94} (\bibinfo {year}
  {1995})},\ \bibinfo {note} {the Physics of Highly Charged Ions}\BibitemShut
  {NoStop}%
\bibitem [{\citenamefont {Jaegl\'{e}}(2006)}]{Jaegle_2006}%
  \BibitemOpen
  \bibfield  {author} {\bibinfo {author} {\bibfnamefont {P.}~\bibnamefont
  {Jaegl\'{e}}},\ }\href@noop {} {\emph {\bibinfo {title} {Coherent sources of
  XUV radiation: Soft X-ray lasers and high-order harmonic generation}}}\
  (\bibinfo  {publisher} {Springer Science+Business Media, Inc},\ \bibinfo
  {year} {2006})\BibitemShut {NoStop}%
\bibitem [{\citenamefont {Silfvast}, \citenamefont {Macklin},\ and\
  \citenamefont {Wood}(1983)}]{Silfvast1983}%
  \BibitemOpen
  \bibfield  {author} {\bibinfo {author} {\bibfnamefont {W.~T.}\ \bibnamefont
  {Silfvast}}, \bibinfo {author} {\bibfnamefont {J.~J.}\ \bibnamefont
  {Macklin}},\ and\ \bibinfo {author} {\bibfnamefont {O.~R.}\ \bibnamefont
  {Wood}},\ }\bibfield  {title} {\enquote {\bibinfo {title} {High -gain
  inner-shell photoionization laser in cd vapor pumped by soft-x-ray radiation
  from a laser-produced plasma source},}\ }\href
  {https://doi.org/10.1364/OL.8.000551} {\bibfield  {journal} {\bibinfo
  {journal} {Opt. Lett.}\ }\textbf {\bibinfo {volume} {8}},\ \bibinfo {pages}
  {551--553} (\bibinfo {year} {1983})}\BibitemShut {NoStop}%
\bibitem [{\citenamefont {Kapteyn}, \citenamefont {Lee},\ and\ \citenamefont
  {Falcone}(1986)}]{Kapteyn1986}%
  \BibitemOpen
  \bibfield  {author} {\bibinfo {author} {\bibfnamefont {H.~C.}\ \bibnamefont
  {Kapteyn}}, \bibinfo {author} {\bibfnamefont {R.~W.}\ \bibnamefont {Lee}},\
  and\ \bibinfo {author} {\bibfnamefont {R.~W.}\ \bibnamefont {Falcone}},\
  }\bibfield  {title} {\enquote {\bibinfo {title} {Observation of a
  short-wavelength laser pumped by auger decay},}\ }\href
  {https://doi.org/10.1103/PhysRevLett.57.2939} {\bibfield  {journal} {\bibinfo
   {journal} {Phys. Rev. Lett.}\ }\textbf {\bibinfo {volume} {57}},\ \bibinfo
  {pages} {2939--2942} (\bibinfo {year} {1986})}\BibitemShut {NoStop}%
\bibitem [{\citenamefont {Pellegrini}(2012)}]{Pellegrini2012}%
  \BibitemOpen
  \bibfield  {author} {\bibinfo {author} {\bibfnamefont {C.}~\bibnamefont
  {Pellegrini}},\ }\bibfield  {title} {\enquote {\bibinfo {title} {The history
  of x-ray free-electron lasers},}\ }\href
  {https://doi.org/10.1140/epjh/e2012-20064-5} {\bibfield  {journal} {\bibinfo
  {journal} {The European Physical Journal H}\ }\textbf {\bibinfo {volume}
  {37}},\ \bibinfo {pages} {659--708} (\bibinfo {year} {2012})}\BibitemShut
  {NoStop}%
\bibitem [{\citenamefont {Kapteyn}(1992)}]{Kapteyn1992}%
  \BibitemOpen
  \bibfield  {author} {\bibinfo {author} {\bibfnamefont {H.~C.}\ \bibnamefont
  {Kapteyn}},\ }\bibfield  {title} {\enquote {\bibinfo {title}
  {Photoionization-pumped x-ray lasers using ultrashort-pulse excitation},}\
  }\href {https://doi.org/10.1364/AO.31.004931} {\bibfield  {journal} {\bibinfo
   {journal} {Appl. Opt.}\ }\textbf {\bibinfo {volume} {31}},\ \bibinfo {pages}
  {4931--4939} (\bibinfo {year} {1992})}\BibitemShut {NoStop}%
\bibitem [{\citenamefont {Rohringer}\ and\ \citenamefont
  {Santra}(2007)}]{Rohringer2007}%
  \BibitemOpen
  \bibfield  {author} {\bibinfo {author} {\bibfnamefont {N.}~\bibnamefont
  {Rohringer}}\ and\ \bibinfo {author} {\bibfnamefont {R.}~\bibnamefont
  {Santra}},\ }\bibfield  {title} {\enquote {\bibinfo {title} {X-ray nonlinear
  optical processes using a self-amplified spontaneous emission free-electron
  laser},}\ }\href {https://doi.org/10.1103/PhysRevA.76.033416} {\bibfield
  {journal} {\bibinfo  {journal} {Phys. Rev. A}\ }\textbf {\bibinfo {volume}
  {76}},\ \bibinfo {pages} {033416} (\bibinfo {year} {2007})}\BibitemShut
  {NoStop}%
\bibitem [{\citenamefont {Rohringer}\ and\ \citenamefont
  {London}(2009)}]{Rohringer2009}%
  \BibitemOpen
  \bibfield  {author} {\bibinfo {author} {\bibfnamefont {N.}~\bibnamefont
  {Rohringer}}\ and\ \bibinfo {author} {\bibfnamefont {R.}~\bibnamefont
  {London}},\ }\bibfield  {title} {\enquote {\bibinfo {title} {Atomic
  inner-shell x-ray laser pumped by an x-ray free-electron laser},}\ }\href
  {https://doi.org/10.1103/PhysRevA.80.013809} {\bibfield  {journal} {\bibinfo
  {journal} {Phys. Rev. A}\ }\textbf {\bibinfo {volume} {80}},\ \bibinfo
  {pages} {013809} (\bibinfo {year} {2009})}\BibitemShut {NoStop}%
\bibitem [{\citenamefont {Tono}\ \emph {et~al.}(2013)\citenamefont {Tono},
  \citenamefont {Togashi}, \citenamefont {Inubushi}, \citenamefont {Sato},
  \citenamefont {Katayama}, \citenamefont {Ogawa}, \citenamefont {Ohashi},
  \citenamefont {Kimura}, \citenamefont {Takahashi}, \citenamefont {Takeshita},
  \citenamefont {Tomizawa}, \citenamefont {Goto}, \citenamefont {Ishikawa},\
  and\ \citenamefont {Yabashi}}]{Tono2013}%
  \BibitemOpen
  \bibfield  {author} {\bibinfo {author} {\bibfnamefont {K.}~\bibnamefont
  {Tono}}, \bibinfo {author} {\bibfnamefont {T.}~\bibnamefont {Togashi}},
  \bibinfo {author} {\bibfnamefont {Y.}~\bibnamefont {Inubushi}}, \bibinfo
  {author} {\bibfnamefont {T.}~\bibnamefont {Sato}}, \bibinfo {author}
  {\bibfnamefont {T.}~\bibnamefont {Katayama}}, \bibinfo {author}
  {\bibfnamefont {K.}~\bibnamefont {Ogawa}}, \bibinfo {author} {\bibfnamefont
  {H.}~\bibnamefont {Ohashi}}, \bibinfo {author} {\bibfnamefont
  {H.}~\bibnamefont {Kimura}}, \bibinfo {author} {\bibfnamefont
  {S.}~\bibnamefont {Takahashi}}, \bibinfo {author} {\bibfnamefont
  {K.}~\bibnamefont {Takeshita}}, \bibinfo {author} {\bibfnamefont
  {H.}~\bibnamefont {Tomizawa}}, \bibinfo {author} {\bibfnamefont
  {S.}~\bibnamefont {Goto}}, \bibinfo {author} {\bibfnamefont {T.}~\bibnamefont
  {Ishikawa}},\ and\ \bibinfo {author} {\bibfnamefont {M.}~\bibnamefont
  {Yabashi}},\ }\bibfield  {title} {\enquote {\bibinfo {title} {Beamline,
  experimental stations and photon beam diagnostics for the hard x-ray free
  electron laser of {SACLA}},}\ }\href
  {https://doi.org/10.1088/1367-2630/15/8/083035} {\bibfield  {journal}
  {\bibinfo  {journal} {New Journal of Physics}\ }\textbf {\bibinfo {volume}
  {15}},\ \bibinfo {pages} {083035} (\bibinfo {year} {2013})}\BibitemShut
  {NoStop}%
\bibitem [{\citenamefont {Nilsen}(2016)}]{Nilsen2016}%
  \BibitemOpen
  \bibfield  {author} {\bibinfo {author} {\bibfnamefont {J.}~\bibnamefont
  {Nilsen}},\ }\bibfield  {title} {\enquote {\bibinfo {title} {Modeling the
  gain of inner-shell x-ray laser transitions in neon, argon, and copper driven
  by x-ray free electron laser radiation using photo-ionization and
  photo-excitation processes},}\ }\href
  {https://doi.org/https://doi.org/10.1016/j.mre.2015.12.001} {\bibfield
  {journal} {\bibinfo  {journal} {Matter and Radiation at Extremes}\ }\textbf
  {\bibinfo {volume} {1}},\ \bibinfo {pages} {76--81} (\bibinfo {year}
  {2016})}\BibitemShut {NoStop}%
\bibitem [{\citenamefont {Ciricosta}\ \emph
  {et~al.}(2016{\natexlab{a}})\citenamefont {Ciricosta}, \citenamefont {Vinko},
  \citenamefont {Barbrel}, \citenamefont {Rackstraw}, \citenamefont {Preston},
  \citenamefont {Burian}, \citenamefont {Chalupsk{\textsurd}$\Omega$},
  \citenamefont {Cho}, \citenamefont {Chung}, \citenamefont {Dakovski},
  \citenamefont {Engelhorn}, \citenamefont
  {H{\textsurd}{\textdegree}jkov{\textsurd}{\textdegree}}, \citenamefont
  {Heimann}, \citenamefont {Holmes}, \citenamefont {Juha}, \citenamefont
  {Krzywinski}, \citenamefont {Lee}, \citenamefont {Toleikis}, \citenamefont
  {Turner}, \citenamefont {Zastrau},\ and\ \citenamefont
  {Wark}}]{Ciricosta2016a}%
  \BibitemOpen
  \bibfield  {author} {\bibinfo {author} {\bibfnamefont {O.}~\bibnamefont
  {Ciricosta}}, \bibinfo {author} {\bibfnamefont {S.~M.}\ \bibnamefont
  {Vinko}}, \bibinfo {author} {\bibfnamefont {B.}~\bibnamefont {Barbrel}},
  \bibinfo {author} {\bibfnamefont {D.~S.}\ \bibnamefont {Rackstraw}}, \bibinfo
  {author} {\bibfnamefont {T.~R.}\ \bibnamefont {Preston}}, \bibinfo {author}
  {\bibfnamefont {T.}~\bibnamefont {Burian}}, \bibinfo {author} {\bibfnamefont
  {J.}~\bibnamefont {Chalupsk{\textsurd}$\Omega$}}, \bibinfo {author}
  {\bibfnamefont {B.~I.}\ \bibnamefont {Cho}}, \bibinfo {author} {\bibfnamefont
  {H.-K.}\ \bibnamefont {Chung}}, \bibinfo {author} {\bibfnamefont {G.~L.}\
  \bibnamefont {Dakovski}}, \bibinfo {author} {\bibfnamefont {K.}~\bibnamefont
  {Engelhorn}}, \bibinfo {author} {\bibfnamefont {V.}~\bibnamefont
  {H{\textsurd}{\textdegree}jkov{\textsurd}{\textdegree}}}, \bibinfo {author}
  {\bibfnamefont {P.}~\bibnamefont {Heimann}}, \bibinfo {author} {\bibfnamefont
  {M.}~\bibnamefont {Holmes}}, \bibinfo {author} {\bibfnamefont
  {L.}~\bibnamefont {Juha}}, \bibinfo {author} {\bibfnamefont {J.}~\bibnamefont
  {Krzywinski}}, \bibinfo {author} {\bibfnamefont {R.~W.}\ \bibnamefont {Lee}},
  \bibinfo {author} {\bibfnamefont {S.}~\bibnamefont {Toleikis}}, \bibinfo
  {author} {\bibfnamefont {J.~J.}\ \bibnamefont {Turner}}, \bibinfo {author}
  {\bibfnamefont {U.}~\bibnamefont {Zastrau}},\ and\ \bibinfo {author}
  {\bibfnamefont {J.~S.}\ \bibnamefont {Wark}},\ }\bibfield  {title} {\enquote
  {\bibinfo {title} {Measurements of continuum lowering in solid-density
  plasmas created from elements and compounds},}\ }\href
  {https://doi.org/10.1038/ncomms11713} {\bibfield  {journal} {\bibinfo
  {journal} {Nature Communications}\ }\textbf {\bibinfo {volume} {7}},\
  \bibinfo {pages} {11713} (\bibinfo {year} {2016}{\natexlab{a}})}\BibitemShut
  {NoStop}%
\bibitem [{\citenamefont {Preston}\ \emph {et~al.}(2017)\citenamefont
  {Preston}, \citenamefont {Vinko}, \citenamefont {Ciricosta}, \citenamefont
  {Hollebon}, \citenamefont {Chung}, \citenamefont {Dakovski}, \citenamefont
  {Krzywinski}, \citenamefont {Minitti}, \citenamefont {Burian}, \citenamefont
  {Chalupsk\'y}, \citenamefont {H\'ajkov\'a}, \citenamefont {Juha},
  \citenamefont {Vozda}, \citenamefont {Zastrau}, \citenamefont {Lee},\ and\
  \citenamefont {Wark}}]{Preston2017}%
  \BibitemOpen
  \bibfield  {author} {\bibinfo {author} {\bibfnamefont {T.~R.}\ \bibnamefont
  {Preston}}, \bibinfo {author} {\bibfnamefont {S.~M.}\ \bibnamefont {Vinko}},
  \bibinfo {author} {\bibfnamefont {O.}~\bibnamefont {Ciricosta}}, \bibinfo
  {author} {\bibfnamefont {P.}~\bibnamefont {Hollebon}}, \bibinfo {author}
  {\bibfnamefont {H.-K.}\ \bibnamefont {Chung}}, \bibinfo {author}
  {\bibfnamefont {G.~L.}\ \bibnamefont {Dakovski}}, \bibinfo {author}
  {\bibfnamefont {J.}~\bibnamefont {Krzywinski}}, \bibinfo {author}
  {\bibfnamefont {M.}~\bibnamefont {Minitti}}, \bibinfo {author} {\bibfnamefont
  {T.}~\bibnamefont {Burian}}, \bibinfo {author} {\bibfnamefont
  {J.}~\bibnamefont {Chalupsk\'y}}, \bibinfo {author} {\bibfnamefont
  {V.}~\bibnamefont {H\'ajkov\'a}}, \bibinfo {author} {\bibfnamefont
  {L.}~\bibnamefont {Juha}}, \bibinfo {author} {\bibfnamefont {V.}~\bibnamefont
  {Vozda}}, \bibinfo {author} {\bibfnamefont {U.}~\bibnamefont {Zastrau}},
  \bibinfo {author} {\bibfnamefont {R.~W.}\ \bibnamefont {Lee}},\ and\ \bibinfo
  {author} {\bibfnamefont {J.~S.}\ \bibnamefont {Wark}},\ }\bibfield  {title}
  {\enquote {\bibinfo {title} {Measurements of the $k$-shell opacity of a
  solid-density magnesium plasma heated by an x-ray free-electron laser},}\
  }\href {https://doi.org/10.1103/PhysRevLett.119.085001} {\bibfield  {journal}
  {\bibinfo  {journal} {Phys. Rev. Lett.}\ }\textbf {\bibinfo {volume} {119}},\
  \bibinfo {pages} {085001} (\bibinfo {year} {2017})}\BibitemShut {NoStop}%
\bibitem [{\citenamefont {van~den Berg}\ \emph
  {et~al.}(2018{\natexlab{a}})\citenamefont {van~den Berg}, \citenamefont
  {Fernandez-Tello}, \citenamefont {Burian}, \citenamefont {Chalupsk\'y},
  \citenamefont {Chung}, \citenamefont {Ciricosta}, \citenamefont {Dakovski},
  \citenamefont {H\'ajkov\'a}, \citenamefont {Hollebon}, \citenamefont {Juha},
  \citenamefont {Krzywinski}, \citenamefont {Lee}, \citenamefont {Minitti},
  \citenamefont {Preston}, \citenamefont {de~la Varga}, \citenamefont {Vozda},
  \citenamefont {Zastrau}, \citenamefont {Wark}, \citenamefont {Velarde},\ and\
  \citenamefont {Vinko}}]{Vandenberg2018}%
  \BibitemOpen
  \bibfield  {author} {\bibinfo {author} {\bibfnamefont {Q.~Y.}\ \bibnamefont
  {van~den Berg}}, \bibinfo {author} {\bibfnamefont {E.~V.}\ \bibnamefont
  {Fernandez-Tello}}, \bibinfo {author} {\bibfnamefont {T.}~\bibnamefont
  {Burian}}, \bibinfo {author} {\bibfnamefont {J.}~\bibnamefont {Chalupsk\'y}},
  \bibinfo {author} {\bibfnamefont {H.-K.}\ \bibnamefont {Chung}}, \bibinfo
  {author} {\bibfnamefont {O.}~\bibnamefont {Ciricosta}}, \bibinfo {author}
  {\bibfnamefont {G.~L.}\ \bibnamefont {Dakovski}}, \bibinfo {author}
  {\bibfnamefont {V.}~\bibnamefont {H\'ajkov\'a}}, \bibinfo {author}
  {\bibfnamefont {P.}~\bibnamefont {Hollebon}}, \bibinfo {author}
  {\bibfnamefont {L.}~\bibnamefont {Juha}}, \bibinfo {author} {\bibfnamefont
  {J.}~\bibnamefont {Krzywinski}}, \bibinfo {author} {\bibfnamefont {R.~W.}\
  \bibnamefont {Lee}}, \bibinfo {author} {\bibfnamefont {M.~P.}\ \bibnamefont
  {Minitti}}, \bibinfo {author} {\bibfnamefont {T.~R.}\ \bibnamefont
  {Preston}}, \bibinfo {author} {\bibfnamefont {A.~G.}\ \bibnamefont {de~la
  Varga}}, \bibinfo {author} {\bibfnamefont {V.}~\bibnamefont {Vozda}},
  \bibinfo {author} {\bibfnamefont {U.}~\bibnamefont {Zastrau}}, \bibinfo
  {author} {\bibfnamefont {J.~S.}\ \bibnamefont {Wark}}, \bibinfo {author}
  {\bibfnamefont {P.}~\bibnamefont {Velarde}},\ and\ \bibinfo {author}
  {\bibfnamefont {S.~M.}\ \bibnamefont {Vinko}},\ }\bibfield  {title} {\enquote
  {\bibinfo {title} {Clocking femtosecond collisional dynamics via resonant
  x-ray spectroscopy},}\ }\href
  {https://doi.org/10.1103/PhysRevLett.120.055002} {\bibfield  {journal}
  {\bibinfo  {journal} {Phys. Rev. Lett.}\ }\textbf {\bibinfo {volume} {120}},\
  \bibinfo {pages} {055002} (\bibinfo {year} {2018}{\natexlab{a}})}\BibitemShut
  {NoStop}%
\bibitem [{\citenamefont {Ciricosta}\ \emph
  {et~al.}(2016{\natexlab{b}})\citenamefont {Ciricosta}, \citenamefont {Vinko},
  \citenamefont {Chung}, \citenamefont {Jackson}, \citenamefont {Lee},
  \citenamefont {Preston}, \citenamefont {Rackstraw},\ and\ \citenamefont
  {Wark}}]{Ciricosta2016}%
  \BibitemOpen
  \bibfield  {author} {\bibinfo {author} {\bibfnamefont {O.}~\bibnamefont
  {Ciricosta}}, \bibinfo {author} {\bibfnamefont {S.~M.}\ \bibnamefont
  {Vinko}}, \bibinfo {author} {\bibfnamefont {H.-K.}\ \bibnamefont {Chung}},
  \bibinfo {author} {\bibfnamefont {C.}~\bibnamefont {Jackson}}, \bibinfo
  {author} {\bibfnamefont {R.~W.}\ \bibnamefont {Lee}}, \bibinfo {author}
  {\bibfnamefont {T.~R.}\ \bibnamefont {Preston}}, \bibinfo {author}
  {\bibfnamefont {D.~S.}\ \bibnamefont {Rackstraw}},\ and\ \bibinfo {author}
  {\bibfnamefont {J.~S.}\ \bibnamefont {Wark}},\ }\bibfield  {title} {\enquote
  {\bibinfo {title} {Detailed model for hot-dense aluminum plasmas generated by
  an x-ray free electron laser},}\ }\href {https://doi.org/10.1063/1.4942540}
  {\bibfield  {journal} {\bibinfo  {journal} {Physics of Plasmas}\ }\textbf
  {\bibinfo {volume} {23}},\ \bibinfo {pages} {022707} (\bibinfo {year}
  {2016}{\natexlab{b}})},\ \Eprint
  {https://arxiv.org/abs/https://doi.org/10.1063/1.4942540}
  {https://doi.org/10.1063/1.4942540} \BibitemShut {NoStop}%
\bibitem [{\citenamefont {Chung}\ \emph {et~al.}(2005)\citenamefont {Chung},
  \citenamefont {Chen}, \citenamefont {Morgan}, \citenamefont {Ralchenko},\
  and\ \citenamefont {Lee}}]{Chung2005}%
  \BibitemOpen
  \bibfield  {author} {\bibinfo {author} {\bibfnamefont {H.-K.}\ \bibnamefont
  {Chung}}, \bibinfo {author} {\bibfnamefont {M.}~\bibnamefont {Chen}},
  \bibinfo {author} {\bibfnamefont {W.}~\bibnamefont {Morgan}}, \bibinfo
  {author} {\bibfnamefont {Y.}~\bibnamefont {Ralchenko}},\ and\ \bibinfo
  {author} {\bibfnamefont {R.}~\bibnamefont {Lee}},\ }\bibfield  {title}
  {\enquote {\bibinfo {title} {Flychk: Generalized population kinetics and
  spectral model for rapid spectroscopic analysis for all elements},}\ }\href
  {https://doi.org/https://doi.org/10.1016/j.hedp.2005.07.001} {\bibfield
  {journal} {\bibinfo  {journal} {High Energy Density Physics}\ }\textbf
  {\bibinfo {volume} {1}},\ \bibinfo {pages} {3--12} (\bibinfo {year}
  {2005})}\BibitemShut {NoStop}%
\bibitem [{\citenamefont {Chung}, \citenamefont {Chen},\ and\ \citenamefont
  {Lee}(2007)}]{Chung2007}%
  \BibitemOpen
  \bibfield  {author} {\bibinfo {author} {\bibfnamefont {H.-K.}\ \bibnamefont
  {Chung}}, \bibinfo {author} {\bibfnamefont {M.}~\bibnamefont {Chen}},\ and\
  \bibinfo {author} {\bibfnamefont {R.}~\bibnamefont {Lee}},\ }\bibfield
  {title} {\enquote {\bibinfo {title} {Extension of atomic configuration sets
  of the non-lte model in the application to the k$\alpha$ diagnostics of hot
  dense matter},}\ }\href
  {https://doi.org/https://doi.org/10.1016/j.hedp.2007.02.001} {\bibfield
  {journal} {\bibinfo  {journal} {High Energy Density Physics}\ }\textbf
  {\bibinfo {volume} {3}},\ \bibinfo {pages} {57--64} (\bibinfo {year}
  {2007})},\ \bibinfo {note} {radiative Properties of Hot Dense
  Matter}\BibitemShut {NoStop}%
\bibitem [{\citenamefont {Ng}\ \emph {et~al.}(1995)\citenamefont {Ng},
  \citenamefont {Celliers}, \citenamefont {Xu},\ and\ \citenamefont
  {Forsman}}]{Ng1995}%
  \BibitemOpen
  \bibfield  {author} {\bibinfo {author} {\bibfnamefont {A.}~\bibnamefont
  {Ng}}, \bibinfo {author} {\bibfnamefont {P.}~\bibnamefont {Celliers}},
  \bibinfo {author} {\bibfnamefont {G.}~\bibnamefont {Xu}},\ and\ \bibinfo
  {author} {\bibfnamefont {A.}~\bibnamefont {Forsman}},\ }\bibfield  {title}
  {\enquote {\bibinfo {title} {Electron-ion equilibration in a strongly coupled
  plasma},}\ }\href {https://doi.org/10.1103/PhysRevE.52.4299} {\bibfield
  {journal} {\bibinfo  {journal} {Phys. Rev. E}\ }\textbf {\bibinfo {volume}
  {52}},\ \bibinfo {pages} {4299--4310} (\bibinfo {year} {1995})}\BibitemShut
  {NoStop}%
\bibitem [{\citenamefont {Nicoul}\ \emph {et~al.}(2011)\citenamefont {Nicoul},
  \citenamefont {Shymanovich}, \citenamefont {Tarasevitch}, \citenamefont
  {von~der Linde},\ and\ \citenamefont {Sokolowski-Tinten}}]{Matthieu2011}%
  \BibitemOpen
  \bibfield  {author} {\bibinfo {author} {\bibfnamefont {M.}~\bibnamefont
  {Nicoul}}, \bibinfo {author} {\bibfnamefont {U.}~\bibnamefont {Shymanovich}},
  \bibinfo {author} {\bibfnamefont {A.}~\bibnamefont {Tarasevitch}}, \bibinfo
  {author} {\bibfnamefont {D.}~\bibnamefont {von~der Linde}},\ and\ \bibinfo
  {author} {\bibfnamefont {K.}~\bibnamefont {Sokolowski-Tinten}},\ }\bibfield
  {title} {\enquote {\bibinfo {title} {Picosecond acoustic response of a
  laser-heated gold-film studied with time-resolved x-ray diffraction},}\
  }\href {https://doi.org/10.1063/1.3584864} {\bibfield  {journal} {\bibinfo
  {journal} {Applied Physics Letters}\ }\textbf {\bibinfo {volume} {98}},\
  \bibinfo {pages} {191902} (\bibinfo {year} {2011})},\ \Eprint
  {https://arxiv.org/abs/https://doi.org/10.1063/1.3584864}
  {https://doi.org/10.1063/1.3584864} \BibitemShut {NoStop}%
\bibitem [{\citenamefont {White}\ \emph {et~al.}(2014)\citenamefont {White},
  \citenamefont {Mabey}, \citenamefont {Gericke}, \citenamefont {Hartley},
  \citenamefont {Doyle}, \citenamefont {McGonegle}, \citenamefont {Rackstraw},
  \citenamefont {Higginbotham},\ and\ \citenamefont {Gregori}}]{White2014}%
  \BibitemOpen
  \bibfield  {author} {\bibinfo {author} {\bibfnamefont {T.~G.}\ \bibnamefont
  {White}}, \bibinfo {author} {\bibfnamefont {P.}~\bibnamefont {Mabey}},
  \bibinfo {author} {\bibfnamefont {D.~O.}\ \bibnamefont {Gericke}}, \bibinfo
  {author} {\bibfnamefont {N.~J.}\ \bibnamefont {Hartley}}, \bibinfo {author}
  {\bibfnamefont {H.~W.}\ \bibnamefont {Doyle}}, \bibinfo {author}
  {\bibfnamefont {D.}~\bibnamefont {McGonegle}}, \bibinfo {author}
  {\bibfnamefont {D.~S.}\ \bibnamefont {Rackstraw}}, \bibinfo {author}
  {\bibfnamefont {A.}~\bibnamefont {Higginbotham}},\ and\ \bibinfo {author}
  {\bibfnamefont {G.}~\bibnamefont {Gregori}},\ }\bibfield  {title} {\enquote
  {\bibinfo {title} {Electron-phonon equilibration in laser-heated gold
  films},}\ }\href {https://doi.org/10.1103/PhysRevB.90.014305} {\bibfield
  {journal} {\bibinfo  {journal} {Phys. Rev. B}\ }\textbf {\bibinfo {volume}
  {90}},\ \bibinfo {pages} {014305} (\bibinfo {year} {2014})}\BibitemShut
  {NoStop}%
\bibitem [{\citenamefont {Ecker}\ and\ \citenamefont
  {Kr{\"o}ll}(1963)}]{Ecker1963}%
  \BibitemOpen
  \bibfield  {author} {\bibinfo {author} {\bibfnamefont {G.}~\bibnamefont
  {Ecker}}\ and\ \bibinfo {author} {\bibfnamefont {W.}~\bibnamefont
  {Kr{\"o}ll}},\ }\bibfield  {title} {\enquote {\bibinfo {title} {Lowering of
  the ionization energy for a plasma in thermodynamic equilibrium},}\
  }\href@noop {} {\bibfield  {journal} {\bibinfo  {journal} {The Physics of
  Fluids}\ }\textbf {\bibinfo {volume} {6}},\ \bibinfo {pages} {62--69}
  (\bibinfo {year} {1963})}\BibitemShut {NoStop}%
\bibitem [{\citenamefont {Ciricosta}\ \emph {et~al.}(2012)\citenamefont
  {Ciricosta}, \citenamefont {Vinko}, \citenamefont {Chung}, \citenamefont
  {Cho}, \citenamefont {Brown}, \citenamefont {Burian}, \citenamefont
  {Chalupsk\'y}, \citenamefont {Engelhorn}, \citenamefont {Falcone},
  \citenamefont {Graves}, \citenamefont {H\'ajkov\'a}, \citenamefont
  {Higginbotham}, \citenamefont {Juha}, \citenamefont {Krzywinski},
  \citenamefont {Lee}, \citenamefont {Messerschmidt}, \citenamefont {Murphy},
  \citenamefont {Ping}, \citenamefont {Rackstraw}, \citenamefont {Scherz},
  \citenamefont {Schlotter}, \citenamefont {Toleikis}, \citenamefont {Turner},
  \citenamefont {Vysin}, \citenamefont {Wang}, \citenamefont {Wu},
  \citenamefont {Zastrau}, \citenamefont {Zhu}, \citenamefont {Lee},
  \citenamefont {Heimann}, \citenamefont {Nagler},\ and\ \citenamefont
  {Wark}}]{Ciricosta2012}%
  \BibitemOpen
  \bibfield  {author} {\bibinfo {author} {\bibfnamefont {O.}~\bibnamefont
  {Ciricosta}}, \bibinfo {author} {\bibfnamefont {S.~M.}\ \bibnamefont
  {Vinko}}, \bibinfo {author} {\bibfnamefont {H.-K.}\ \bibnamefont {Chung}},
  \bibinfo {author} {\bibfnamefont {B.-I.}\ \bibnamefont {Cho}}, \bibinfo
  {author} {\bibfnamefont {C.~R.~D.}\ \bibnamefont {Brown}}, \bibinfo {author}
  {\bibfnamefont {T.}~\bibnamefont {Burian}}, \bibinfo {author} {\bibfnamefont
  {J.}~\bibnamefont {Chalupsk\'y}}, \bibinfo {author} {\bibfnamefont
  {K.}~\bibnamefont {Engelhorn}}, \bibinfo {author} {\bibfnamefont {R.~W.}\
  \bibnamefont {Falcone}}, \bibinfo {author} {\bibfnamefont {C.}~\bibnamefont
  {Graves}}, \bibinfo {author} {\bibfnamefont {V.}~\bibnamefont {H\'ajkov\'a}},
  \bibinfo {author} {\bibfnamefont {A.}~\bibnamefont {Higginbotham}}, \bibinfo
  {author} {\bibfnamefont {L.}~\bibnamefont {Juha}}, \bibinfo {author}
  {\bibfnamefont {J.}~\bibnamefont {Krzywinski}}, \bibinfo {author}
  {\bibfnamefont {H.~J.}\ \bibnamefont {Lee}}, \bibinfo {author} {\bibfnamefont
  {M.}~\bibnamefont {Messerschmidt}}, \bibinfo {author} {\bibfnamefont {C.~D.}\
  \bibnamefont {Murphy}}, \bibinfo {author} {\bibfnamefont {Y.}~\bibnamefont
  {Ping}}, \bibinfo {author} {\bibfnamefont {D.~S.}\ \bibnamefont {Rackstraw}},
  \bibinfo {author} {\bibfnamefont {A.}~\bibnamefont {Scherz}}, \bibinfo
  {author} {\bibfnamefont {W.}~\bibnamefont {Schlotter}}, \bibinfo {author}
  {\bibfnamefont {S.}~\bibnamefont {Toleikis}}, \bibinfo {author}
  {\bibfnamefont {J.~J.}\ \bibnamefont {Turner}}, \bibinfo {author}
  {\bibfnamefont {L.}~\bibnamefont {Vysin}}, \bibinfo {author} {\bibfnamefont
  {T.}~\bibnamefont {Wang}}, \bibinfo {author} {\bibfnamefont {B.}~\bibnamefont
  {Wu}}, \bibinfo {author} {\bibfnamefont {U.}~\bibnamefont {Zastrau}},
  \bibinfo {author} {\bibfnamefont {D.}~\bibnamefont {Zhu}}, \bibinfo {author}
  {\bibfnamefont {R.~W.}\ \bibnamefont {Lee}}, \bibinfo {author} {\bibfnamefont
  {P.}~\bibnamefont {Heimann}}, \bibinfo {author} {\bibfnamefont
  {B.}~\bibnamefont {Nagler}},\ and\ \bibinfo {author} {\bibfnamefont {J.~S.}\
  \bibnamefont {Wark}},\ }\bibfield  {title} {\enquote {\bibinfo {title}
  {Direct measurements of the ionization potential depression in a dense
  plasma},}\ }\href {https://doi.org/10.1103/PhysRevLett.109.065002} {\bibfield
   {journal} {\bibinfo  {journal} {Phys. Rev. Lett.}\ }\textbf {\bibinfo
  {volume} {109}},\ \bibinfo {pages} {065002} (\bibinfo {year}
  {2012})}\BibitemShut {NoStop}%
\bibitem [{\citenamefont {Young}\ \emph {et~al.}(2010)\citenamefont {Young},
  \citenamefont {Kanter}, \citenamefont {Kr\"{a}ssig}, \citenamefont {Li},
  \citenamefont {March}, \citenamefont {Pratt}, \citenamefont {Santra},
  \citenamefont {Southworth}, \citenamefont {Rohringer}, \citenamefont
  {DiMauro}, \citenamefont {Doumy}, \citenamefont {Roedig}, \citenamefont
  {Berrah}, \citenamefont {Fang}, \citenamefont {Hoener}, \citenamefont
  {Bucksbaum}, \citenamefont {Cryan}, \citenamefont {Ghimire}, \citenamefont
  {Glownia}, \citenamefont {Reis}, \citenamefont {Bozek}, \citenamefont
  {Bostedt},\ and\ \citenamefont {Messerschmidt}}]{Young2010}%
  \BibitemOpen
  \bibfield  {author} {\bibinfo {author} {\bibfnamefont {L.}~\bibnamefont
  {Young}}, \bibinfo {author} {\bibfnamefont {E.~P.}\ \bibnamefont {Kanter}},
  \bibinfo {author} {\bibfnamefont {B.}~\bibnamefont {Kr\"{a}ssig}}, \bibinfo
  {author} {\bibfnamefont {Y.}~\bibnamefont {Li}}, \bibinfo {author}
  {\bibfnamefont {A.~M.}\ \bibnamefont {March}}, \bibinfo {author}
  {\bibfnamefont {S.~T.}\ \bibnamefont {Pratt}}, \bibinfo {author}
  {\bibfnamefont {R.}~\bibnamefont {Santra}}, \bibinfo {author} {\bibfnamefont
  {S.~H.}\ \bibnamefont {Southworth}}, \bibinfo {author} {\bibfnamefont
  {N.}~\bibnamefont {Rohringer}}, \bibinfo {author} {\bibfnamefont {L.~F.}\
  \bibnamefont {DiMauro}}, \bibinfo {author} {\bibfnamefont {G.}~\bibnamefont
  {Doumy}}, \bibinfo {author} {\bibfnamefont {C.~A.}\ \bibnamefont {Roedig}},
  \bibinfo {author} {\bibfnamefont {N.}~\bibnamefont {Berrah}}, \bibinfo
  {author} {\bibfnamefont {L.}~\bibnamefont {Fang}}, \bibinfo {author}
  {\bibfnamefont {M.}~\bibnamefont {Hoener}}, \bibinfo {author} {\bibfnamefont
  {P.~H.}\ \bibnamefont {Bucksbaum}}, \bibinfo {author} {\bibfnamefont {J.~P.}\
  \bibnamefont {Cryan}}, \bibinfo {author} {\bibfnamefont {S.}~\bibnamefont
  {Ghimire}}, \bibinfo {author} {\bibfnamefont {J.~M.}\ \bibnamefont
  {Glownia}}, \bibinfo {author} {\bibfnamefont {D.~A.}\ \bibnamefont {Reis}},
  \bibinfo {author} {\bibfnamefont {J.~D.}\ \bibnamefont {Bozek}}, \bibinfo
  {author} {\bibfnamefont {C.}~\bibnamefont {Bostedt}},\ and\ \bibinfo {author}
  {\bibfnamefont {M.}~\bibnamefont {Messerschmidt}},\ }\bibfield  {title}
  {\enquote {\bibinfo {title} {Femtosecond electronic response of atoms to
  ultra-intense x-rays},}\ }\href {https://doi.org/10.1038/nature09177}
  {\bibfield  {journal} {\bibinfo  {journal} {Nature}\ }\textbf {\bibinfo
  {volume} {466}},\ \bibinfo {pages} {56--61} (\bibinfo {year}
  {2010})}\BibitemShut {NoStop}%
\bibitem [{\citenamefont {Ciricosta}\ \emph {et~al.}(2011)\citenamefont
  {Ciricosta}, \citenamefont {Chung}, \citenamefont {Lee},\ and\ \citenamefont
  {Wark}}]{Ciricosta2011}%
  \BibitemOpen
  \bibfield  {author} {\bibinfo {author} {\bibfnamefont {O.}~\bibnamefont
  {Ciricosta}}, \bibinfo {author} {\bibfnamefont {H.-K.}\ \bibnamefont
  {Chung}}, \bibinfo {author} {\bibfnamefont {R.~W.}\ \bibnamefont {Lee}},\
  and\ \bibinfo {author} {\bibfnamefont {J.~S.}\ \bibnamefont {Wark}},\
  }\bibfield  {title} {\enquote {\bibinfo {title} {Simulations of neon
  irradiated by intense x-ray laser radiation},}\ }\href
  {https://doi.org/https://doi.org/10.1016/j.hedp.2011.02.003} {\bibfield
  {journal} {\bibinfo  {journal} {High Energy Density Physics}\ }\textbf
  {\bibinfo {volume} {7}},\ \bibinfo {pages} {111--116} (\bibinfo {year}
  {2011})}\BibitemShut {NoStop}%
\bibitem [{\citenamefont {Vinko}\ \emph {et~al.}(2012)\citenamefont {Vinko},
  \citenamefont {Ciricosta}, \citenamefont {Cho}, \citenamefont {Engelhorn},
  \citenamefont {Chung}, \citenamefont {Brown}, \citenamefont {Burian},
  \citenamefont {Chalupsk{\'y}}, \citenamefont {Falcone}, \citenamefont
  {Graves}, \citenamefont {H{\'a}jkov{\'a}}, \citenamefont {Higginbotham},
  \citenamefont {Juha}, \citenamefont {Krzywinski}, \citenamefont {Lee},
  \citenamefont {Messerschmidt}, \citenamefont {Murphy}, \citenamefont {Ping},
  \citenamefont {Scherz}, \citenamefont {Schlotter}, \citenamefont {Toleikis},
  \citenamefont {Turner}, \citenamefont {Vysin}, \citenamefont {Wang},
  \citenamefont {Wu}, \citenamefont {Zastrau}, \citenamefont {Zhu},
  \citenamefont {Lee}, \citenamefont {Heimann}, \citenamefont {Nagler},\ and\
  \citenamefont {Wark}}]{Vinko2012}%
  \BibitemOpen
  \bibfield  {author} {\bibinfo {author} {\bibfnamefont {S.~M.}\ \bibnamefont
  {Vinko}}, \bibinfo {author} {\bibfnamefont {O.}~\bibnamefont {Ciricosta}},
  \bibinfo {author} {\bibfnamefont {B.~I.}\ \bibnamefont {Cho}}, \bibinfo
  {author} {\bibfnamefont {K.}~\bibnamefont {Engelhorn}}, \bibinfo {author}
  {\bibfnamefont {H.-K.}\ \bibnamefont {Chung}}, \bibinfo {author}
  {\bibfnamefont {C.~R.~D.}\ \bibnamefont {Brown}}, \bibinfo {author}
  {\bibfnamefont {T.}~\bibnamefont {Burian}}, \bibinfo {author} {\bibfnamefont
  {J.}~\bibnamefont {Chalupsk{\'y}}}, \bibinfo {author} {\bibfnamefont {R.~W.}\
  \bibnamefont {Falcone}}, \bibinfo {author} {\bibfnamefont {C.}~\bibnamefont
  {Graves}}, \bibinfo {author} {\bibfnamefont {V.}~\bibnamefont
  {H{\'a}jkov{\'a}}}, \bibinfo {author} {\bibfnamefont {A.}~\bibnamefont
  {Higginbotham}}, \bibinfo {author} {\bibfnamefont {L.}~\bibnamefont {Juha}},
  \bibinfo {author} {\bibfnamefont {J.}~\bibnamefont {Krzywinski}}, \bibinfo
  {author} {\bibfnamefont {H.~J.}\ \bibnamefont {Lee}}, \bibinfo {author}
  {\bibfnamefont {M.}~\bibnamefont {Messerschmidt}}, \bibinfo {author}
  {\bibfnamefont {C.~D.}\ \bibnamefont {Murphy}}, \bibinfo {author}
  {\bibfnamefont {Y.}~\bibnamefont {Ping}}, \bibinfo {author} {\bibfnamefont
  {A.}~\bibnamefont {Scherz}}, \bibinfo {author} {\bibfnamefont
  {W.}~\bibnamefont {Schlotter}}, \bibinfo {author} {\bibfnamefont
  {S.}~\bibnamefont {Toleikis}}, \bibinfo {author} {\bibfnamefont {J.~J.}\
  \bibnamefont {Turner}}, \bibinfo {author} {\bibfnamefont {L.}~\bibnamefont
  {Vysin}}, \bibinfo {author} {\bibfnamefont {T.}~\bibnamefont {Wang}},
  \bibinfo {author} {\bibfnamefont {B.}~\bibnamefont {Wu}}, \bibinfo {author}
  {\bibfnamefont {U.}~\bibnamefont {Zastrau}}, \bibinfo {author} {\bibfnamefont
  {D.}~\bibnamefont {Zhu}}, \bibinfo {author} {\bibfnamefont {R.~W.}\
  \bibnamefont {Lee}}, \bibinfo {author} {\bibfnamefont {P.~A.}\ \bibnamefont
  {Heimann}}, \bibinfo {author} {\bibfnamefont {B.}~\bibnamefont {Nagler}},\
  and\ \bibinfo {author} {\bibfnamefont {J.~S.}\ \bibnamefont {Wark}},\
  }\bibfield  {title} {\enquote {\bibinfo {title} {Creation and diagnosis of a
  solid-density plasma with an x-ray free-electron laser},}\ }\href
  {https://doi.org/10.1038/nature10746} {\bibfield  {journal} {\bibinfo
  {journal} {Nature}\ }\textbf {\bibinfo {volume} {482}},\ \bibinfo {pages}
  {59--62} (\bibinfo {year} {2012})}\BibitemShut {NoStop}%
\bibitem [{\citenamefont {Vinko}\ \emph {et~al.}(2015)\citenamefont {Vinko},
  \citenamefont {Ciricosta}, \citenamefont {Preston}, \citenamefont
  {Rackstraw}, \citenamefont {Brown}, \citenamefont {Burian}, \citenamefont
  {Chalupsk{\'y}}, \citenamefont {Cho}, \citenamefont {Chung}, \citenamefont
  {Engelhorn}, \citenamefont {Falcone}, \citenamefont {Fiokovinini},
  \citenamefont {H{\'a}jkov{\'a}}, \citenamefont {Heimann}, \citenamefont
  {Juha}, \citenamefont {Lee}, \citenamefont {Lee}, \citenamefont
  {Messerschmidt}, \citenamefont {Nagler}, \citenamefont {Schlotter},
  \citenamefont {Turner}, \citenamefont {Vysin}, \citenamefont {Zastrau},\ and\
  \citenamefont {Wark}}]{Vinko2015}%
  \BibitemOpen
  \bibfield  {author} {\bibinfo {author} {\bibfnamefont {S.~M.}\ \bibnamefont
  {Vinko}}, \bibinfo {author} {\bibfnamefont {O.}~\bibnamefont {Ciricosta}},
  \bibinfo {author} {\bibfnamefont {T.~R.}\ \bibnamefont {Preston}}, \bibinfo
  {author} {\bibfnamefont {D.~S.}\ \bibnamefont {Rackstraw}}, \bibinfo {author}
  {\bibfnamefont {C.~R.~D.}\ \bibnamefont {Brown}}, \bibinfo {author}
  {\bibfnamefont {T.}~\bibnamefont {Burian}}, \bibinfo {author} {\bibfnamefont
  {J.}~\bibnamefont {Chalupsk{\'y}}}, \bibinfo {author} {\bibfnamefont {B.~I.}\
  \bibnamefont {Cho}}, \bibinfo {author} {\bibfnamefont {H.-K.}\ \bibnamefont
  {Chung}}, \bibinfo {author} {\bibfnamefont {K.}~\bibnamefont {Engelhorn}},
  \bibinfo {author} {\bibfnamefont {R.~W.}\ \bibnamefont {Falcone}}, \bibinfo
  {author} {\bibfnamefont {R.}~\bibnamefont {Fiokovinini}}, \bibinfo {author}
  {\bibfnamefont {V.}~\bibnamefont {H{\'a}jkov{\'a}}}, \bibinfo {author}
  {\bibfnamefont {P.~A.}\ \bibnamefont {Heimann}}, \bibinfo {author}
  {\bibfnamefont {L.}~\bibnamefont {Juha}}, \bibinfo {author} {\bibfnamefont
  {H.~J.}\ \bibnamefont {Lee}}, \bibinfo {author} {\bibfnamefont {R.~W.}\
  \bibnamefont {Lee}}, \bibinfo {author} {\bibfnamefont {M.}~\bibnamefont
  {Messerschmidt}}, \bibinfo {author} {\bibfnamefont {B.}~\bibnamefont
  {Nagler}}, \bibinfo {author} {\bibfnamefont {W.}~\bibnamefont {Schlotter}},
  \bibinfo {author} {\bibfnamefont {J.~J.}\ \bibnamefont {Turner}}, \bibinfo
  {author} {\bibfnamefont {L.}~\bibnamefont {Vysin}}, \bibinfo {author}
  {\bibfnamefont {U.}~\bibnamefont {Zastrau}},\ and\ \bibinfo {author}
  {\bibfnamefont {J.~S.}\ \bibnamefont {Wark}},\ }\bibfield  {title} {\enquote
  {\bibinfo {title} {Investigation of femtosecond collisional ionization rates
  in a solid-density aluminium plasma},}\ }\href
  {https://doi.org/10.1038/ncomms7397} {\bibfield  {journal} {\bibinfo
  {journal} {Nature Communications}\ }\textbf {\bibinfo {volume} {6}},\
  \bibinfo {pages} {6397} (\bibinfo {year} {2015})}\BibitemShut {NoStop}%
\bibitem [{\citenamefont {Ren}\ \emph {et~al.}(2022)\citenamefont {Ren},
  \citenamefont {Shi}, \citenamefont {Berg}, \citenamefont {Firmansyah},
  \citenamefont {Chung}, \citenamefont {Fernandez-Tello}, \citenamefont
  {Velarde}, \citenamefont {Wark},\ and\ \citenamefont {Vinko}}]{Shenyuan2022}%
  \BibitemOpen
  \bibfield  {author} {\bibinfo {author} {\bibfnamefont {S.}~\bibnamefont
  {Ren}}, \bibinfo {author} {\bibfnamefont {Y.}~\bibnamefont {Shi}}, \bibinfo
  {author} {\bibfnamefont {Q.~Y. v.~d.}\ \bibnamefont {Berg}}, \bibinfo
  {author} {\bibfnamefont {M.}~\bibnamefont {Firmansyah}}, \bibinfo {author}
  {\bibfnamefont {H.-K.}\ \bibnamefont {Chung}}, \bibinfo {author}
  {\bibfnamefont {E.~V.}\ \bibnamefont {Fernandez-Tello}}, \bibinfo {author}
  {\bibfnamefont {P.}~\bibnamefont {Velarde}}, \bibinfo {author} {\bibfnamefont
  {J.~S.}\ \bibnamefont {Wark}},\ and\ \bibinfo {author} {\bibfnamefont
  {S.~M.}\ \bibnamefont {Vinko}},\ }\href
  {https://doi.org/10.48550/ARXIV.2208.00573} {\enquote {\bibinfo {title}
  {Non-thermal evolution of dense plasmas driven by intense x-ray fields},}\ }
  (\bibinfo {year} {2022})\BibitemShut {NoStop}%
\bibitem [{\citenamefont {van~den Berg}\ \emph
  {et~al.}(2018{\natexlab{b}})\citenamefont {van~den Berg} \emph
  {et~al.}}]{vandenberg:2018}%
  \BibitemOpen
  \bibfield  {author} {\bibinfo {author} {\bibfnamefont {Q.}~\bibnamefont
  {van~den Berg}} \emph {et~al.},\ }\bibfield  {title} {\enquote {\bibinfo
  {title} {Clocking femtosecond collisional dynamics via resonant x-ray
  spectroscopy},}\ }\href@noop {} {\bibfield  {journal} {\bibinfo  {journal}
  {Phys.\ Rev.\ Lett.}\ }\textbf {\bibinfo {volume} {120}},\ \bibinfo {pages}
  {055002} (\bibinfo {year} {2018}{\natexlab{b}})}\BibitemShut {NoStop}%
\bibitem [{\citenamefont {{Lotz}}(1967)}]{1967ZPhy..206..205L}%
  \BibitemOpen
  \bibfield  {author} {\bibinfo {author} {\bibfnamefont {W.}~\bibnamefont
  {{Lotz}}},\ }\bibfield  {title} {\enquote {\bibinfo {title} {{An empirical
  formula for the electron-impact ionization cross-section}},}\ }\href
  {https://doi.org/10.1007/BF01325928} {\bibfield  {journal} {\bibinfo
  {journal} {Zeitschrift fur Physik}\ }\textbf {\bibinfo {volume} {206}},\
  \bibinfo {pages} {205--211} (\bibinfo {year} {1967})}\BibitemShut {NoStop}%
\bibitem [{\citenamefont {Lotz}(1968)}]{lotz:1968}%
  \BibitemOpen
  \bibfield  {author} {\bibinfo {author} {\bibfnamefont {W.}~\bibnamefont
  {Lotz}},\ }\bibfield  {title} {\enquote {\bibinfo {title} {Electron-impact
  ionization cross-sections and ionization rate coefficients for atoms and ions
  from hydrogen to calcium},}\ }\href@noop {} {\bibfield  {journal} {\bibinfo
  {journal} {Zeitschrift Physik}\ }\textbf {\bibinfo {volume} {216}},\ \bibinfo
  {pages} {241} (\bibinfo {year} {1968})}\BibitemShut {NoStop}%
\end{thebibliography}%

\end{document}